\documentclass[12pt, a4paper]{iopart}
\usepackage{graphicx}
\usepackage{cite}
\usepackage{iopams}
\usepackage{bm}
\usepackage{wasysym}
\usepackage[T1]{fontenc}
\begin{document}

\title{Ground state of an ultracold Fermi gas of tilted dipoles in elongated traps}

\author{Vladimir Velji\'{c}$^1$, Aristeu R.~P.~Lima$^2$, Lauriane Chomaz$^3$, Simon Baier$^3$, Manfred J.~Mark$^{3,4}$, Francesca Ferlaino$^{3,4}$, Axel Pelster$^5$ and Antun Bala\v{z}$^1$}
\address{$^1$Scientific Computing Laboratory, Center for the Study of Complex Systems, Institute of Physics Belgrade,
University of Belgrade, Serbia\\
$^2$Universidade da Integra\c{c}\~{a}o Internacional da Lusofonia Afro-Brasileira, Campus das Auroras, Acarape-Cear\'{a}, Brazil\\
$^3$Institute for Experimental Physics, University of Innsbruck, Innsbruck, Austria\\
$^4$Institute for Quantum Optics and Quantum Information, Austrian Academy of Sciences, Innsbruck, Austria\\
$^5$Physics Department and Research Center OPTIMAS, Technische Universit\"at Kaiserslautern, Germany}
\ead{vladimir.veljic@ipb.ac.rs}
 
\begin{abstract}
Many-body dipolar effects in Fermi gases are quite subtle as they energetically compete with the large kinetic energy at and below the Fermi surface (FS). Recently it was experimentally observed in a sample of erbium atoms that its FS is deformed from a sphere to an ellipsoid due to the presence of the anisotropic and long-range dipole-dipole interaction (Aikawa et al 2014 {\em Science} {\bf 345} 1484). Moreover, it was suggested that, when the dipoles are rotated by means of an external field, the Fermi surface follows their rotation, thereby keeping the major axis of the momentum-space ellipsoid parallel to the dipoles. Here we generalise a previous Hartree-Fock mean-field theory to systems confined in an elongated triaxial trap with an arbitrary orientation of the dipoles relative to the trap. With this we study for the first time the effects of the dipoles' arbitrary orientation on the ground-state properties of the system. Furthermore, taking into account the geometry of the system, we show how the ellipsoidal FS deformation can be reconstructed, assuming ballistic expansion, from the experimentally measurable real-space aspect ratio after a free expansion. We compare our theoretical results with new experimental data measured with an erbium Fermi gas for various trap parameters and dipole orientations. The observed remarkable agreement demonstrates the ability of our model to capture the full angular dependence of the FS deformation. Moreover, for systems with even higher dipole moment, our theory predicts an additional unexpected effect: the FS does not simply follow rigidly the orientation of the dipoles, but softens showing a change in the aspect ratio depending on the dipoles' orientation relative to the trap geometry, as well as on the trap anisotropy itself. Our theory provides the basis for understanding and interpreting phenomena in which the investigated physics depends on the underlying structure of the FS, such as fermionic pairing and superfluidity.
\end{abstract}


\section{Introduction}
\label{sec:intro}

Since the first experimental realisation of a dipolar Bose-Einstein condensate (BEC) of chromium atoms \cite{Griesmaier} and the subsequent demonstration of the presence of the anisotropic and long-range dipole-dipole interaction (DDI) in the laboratory \cite{Pfau}, dipolar quantum gases have developed into a vast and fast-growing research field. Indeed, the interplay of the DDI and the isotropic and short-range contact interaction between the particles in these systems makes them particularly intriguing from both the experimental and the theoretical point of view \cite{Baranov,PfauRep,review_zoller}.

More recently, BECs of even more magnetic species, i.e., dysprosium ($ m=10 \mu_{\rm B}$) \cite{Dy-boson} and erbium ($m=7 \mu_{\rm B}$) \cite{Er-boson} have been created. Here, $\mu_{\rm B}$ denotes the Bohr magneton.
Such species exhibit fascinating phenomena, such as the Rosensweig instability \cite{PfauNature}, the emergence of quantum-stabilised droplets \cite{PfauPRL, Francesca2,PfauNature2} and roton quasiparticles \cite{Francesca-roton}.  
Correspondingly, all these developments triggered much theoretical work, including, but not limited to, the numerical effort to simulate dipolar quantum gases in fully anisotropic traps \cite{GP1,GP2,GP3,GP4,GP5}, the roton instability in pancake-shaped condensates \cite{Fedorov1, Fedorov2, Nath}, the investigation of beyond-mean-field effects in one-component \cite{Lima3,Lima4} and two-component \cite{Pastukhov} gases, the formation of the previously observed droplets \cite{Xi,Santos1, Blakie4}, their ground-state properties and elementary excitations \cite{Santos2, Blakie3, Blakie5}, the role of three-body interactions \cite{Blakie2}, and the self-bounded nature of the droplets \cite{Blakie4}.

In parallel, quantum-degenerate dipolar Fermi gases of dysprosium \cite{Dy1}, erbium \cite{Er} and chromium \cite{Cr} became also available in experiments. 
Remarkably, identical fermions of dipolar character do interact even in the low-energy limit because of the peculiar form of the dipolar scattering \cite{sigma}. 
Few-body scattering experiments have indeed confirmed universal scaling in dipolar scattering among fermions \cite{Er, tau, Lev1}. Many-body dipolar effects in Fermi gases are much more subtle to observe because of the competition with the large kinetic energy stored in the Fermi sphere, which leads to the Fermi pressure.
Recently, the key observation of the Fermi surface (FS) deformation was made \cite{Francesca}, confirming previous theoretical predictions \cite{Sogo}.

It is well known that in the case of a single-component Fermi gas, the isotropic and short-range contact interaction is suppressed by the Pauli exclusion principle. Also, we know that the Fermi surface is a sphere, as a consequence of the symmetry of the Pauli pressure. Theoretical predictions which take the DDI into account, however, have shown that the antisymmetry of the wave function leads to the deformation of the Fermi surface into an ellipsoid \cite{Sogo}. The ground-state \cite{Zhang,Baillie} and the dynamic properties of such systems have been systematically investigated theoretically and numerically in the collisionless regime \cite{HeZhang2Yi,Sogo2,Sogo2c}, in the hydrodynamic regime \cite{lima1,lima2}, as well as in the whole collisional range from one limiting case to the other \cite{Veljic1,Falk}. The FS deformation was also recently theoretically studied in mixtures of dipolar and non-dipolar fermions \cite{Baarsma}, as well as in the presence of a weak lattice confinement \cite{Lemeshko}.

Within the Hartree-Fock mean-field theory for a many-body system, the interaction energy of the system is expressed by means of the Hartree direct term, which gives rise to a deformation of the atomic cloud in real space \cite{goral}, and the Fock exchange term, which leads to a deformation of the FS in momentum space \cite{Sogo}.
The Hartree-Fock mean-field approximation, which includes energy terms up to first-order in the DDI, is sufficiently accurate to qualitatively explain and quantitatively describe results of ongoing experiments. However, up to now, existing theories are limited to a fixed orientation of the dipoles, which has to coincide with one of the trap axes \cite{goral,Sogo,Falk,Veljic1}. Such a restriction greatly simplified theoretical considerations, but, on the other hand, limited their scope since the anisotropy of the DDI is best controlled by the dipoles' orientation with respect to the trap axes.

Motivated by this, we develop a general theory to describe the ground state of a dipolar Fermi gas for an arbitrary orientation of the dipoles and trap geometry. Our full theoretical description provides a substantial advance in understanding dipolar phenomena and in describing experimental observations in a very broad parameter range, see e.g.~\cite{Francesca}. Our theory captures both the cloud shape in real space and the FS in momentum space.
To test our theory, we have performed new measurements of the FS deformation in an erbium Fermi gas for various traps and dipole orientations. The comparison between the theory and the experiment shows a remarkable agreement, demonstrating the predictive power of our newly developed theory to calculate the system's behaviour. Moreover, whereas in the Er case the deformed FS rigidly rotates with the dipole orientation, our theory also predicts a softening of the FS in systems with larger DDI. There the degree of deformation also changes depending on the dipoles' orientation.

The approach presented here is very general and can be applied to both fermionic atoms and molecules with electric \cite{KRb, vitanov, NaK1, DSJin} or magnetic \cite{Er2} dipole moments, and any triaxial trap geometry. Our calculation provides a starting point to address more complex dipolar phenomena. Indeed, many physical properties depend on the shape of the FS and on its deformation, as the FS is directly connected to the density distribution in momentum space. For instance, one can revisit a pairing problem within a one-component dipolar Fermi gas, where in a previous work by Baranov et al~\cite{Baranov-pairing} it was assumed that the FS is spherically symmetric. A relevant question for future investigations is whether one can instead have both a deformation of the Fermi sphere and a pairing of fermions at the same time, and if a critical deformation exists for which the pairing mechanism breaks down.

The paper is structured as follows. In section~\ref{sec:theo_mod} we introduce our theoretical model and several suitable ans\"atze for the form of the system's Wigner function. Considering the Hartree-Fock total energy of the system, we identify the ansatz that yields minimal energy for the ground state and use it for all further calculations. In section~\ref{sec:res} we present the generalised theory and our main results for the FS deformation and real-space magnetostriction. In particular, we discuss the behaviour of the variational parameters and their impact upon the Hartree-Fock total energy for the considered system. We also study in detail the ground-state properties for an arbitrary orientation of the dipoles, as well as for different parameters of the system, e.g., trap frequencies, number of particles, and dipolar species. Afterwards, in section~\ref{compa_exp} we directly compare our theoretical predictions with the novel experimental data. Finally, section~\ref{sec:con} gathers our concluding remarks and gives an outlook for future research.

\section{Theoretical model}
\label{sec:theo_mod}

\begin{figure}[!b]
\centering
\includegraphics[width=6cm]{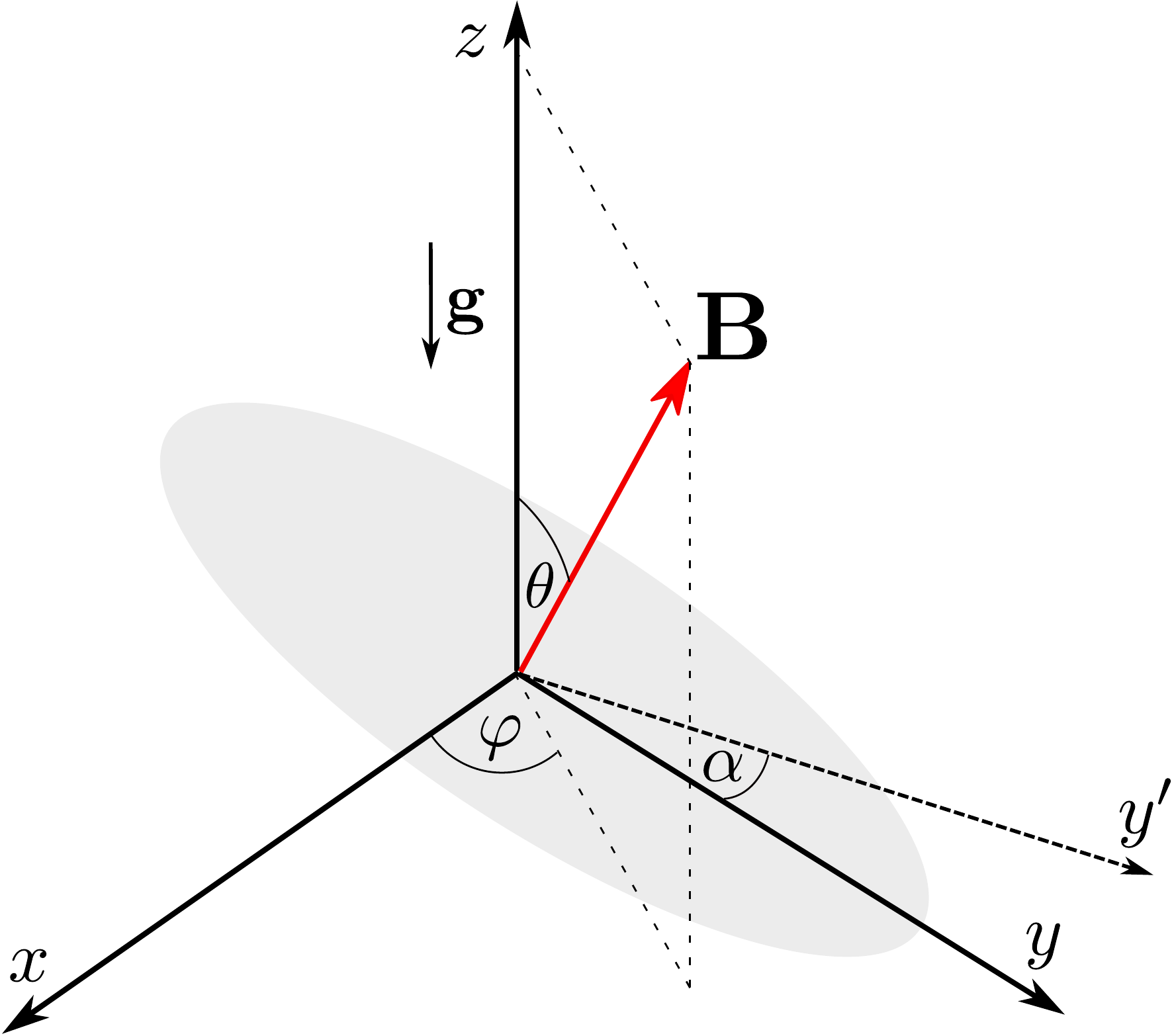}
\caption{Schematic illustration of the most general arbitrary geometry of a dipolar Fermi gas, which corresponds to the one used in the Innsbruck experiment with erbium atoms, see later and also reference \cite{Francesca}. Axes $x, y, z$ indicate the harmonic trap axes. The external magnetic field ${\bf B}=B{\bf e}$ defines the orientation of the atomic dipoles, which is given by the spherical coordinates $\theta$ and $\varphi$. Earth's gravitational field is parallel to the $z$ axis. The imaging axis, denoted by $y'$, lies in the $xy$ plane, and forms an angle $\alpha$ with the $y$ axis.}
\label{fig:exp_setup}
\end{figure}

We consider an ultracold quantum-degenerate dipolar gas at zero temperature consisting of $N$ identical spin-polarized fermions of mass $M$. The fermions have a strong dipolar character, arising from either a large magnetic or an electric dipole moment $\bf m$. Moreover, as is usual in the experiments, we assume that all dipoles are polarised along a single direction, as their orientation can be controlled by an external field, see, e.g., \cite{Francesca, NaK1, TiltingPfau, Francesca2}. To account for the influence of the dipoles' orientation, we will consider the most general possible orientation of an external field, as depicted in figure~\ref{fig:exp_setup}, where, e.g., the magnetic field ${\bf B}= B {\bf e}$ is oriented along the unit vector ${\bf e}$. The trap axes set the reference frame. Additionally, we account for a possible off-axis imaging and consider the case of an imaging beam forming an angle $\alpha$ with the $y$ axis, as shown in figure~\ref{fig:exp_setup}.

Since the Pauli exclusion principle inhibits short-range contact interaction, the long-range DDI between the particles dominates the interaction behaviour of the system. If the polarisation direction of the dipoles is defined by a unit vector $\bf e$, the DDI is described by
\begin{equation}
 V_{\rm dd}({\bf r})=-\frac{C_\mathrm{dd}}{4\pi}\frac{3({\bf r \cdot e})^2-{\bf r}^2}{r^5}\, , 
 \label{eq:ddi}
\end{equation}
where $\bf r$ denotes the relative position of two dipoles and $C_\mathrm{dd}$ is the dipolar interaction strength. For electric dipoles, it is given by $C^{}_\mathrm{dd}=m^2 / \varepsilon_0$ with $\varepsilon_0$ being the vacuum permittivity, while
for magnetic dipoles $C_\mathrm{dd}=\mu_0 m^2$, where $\mu_0$ is the vacuum permeability. In this context, an important role is played by the Fourier transform of the DDI potential \cite{Pfau_Fourier}
\begin{equation}
\tilde{V}_{\rm dd}({\bf k})=\frac{C_\mathrm{dd} }{3}\left[3 \frac{({\bf e \cdot k})^2}{k^2}-1\right]\, ,
\label{eq:FTddi}
\end{equation}
as it simplifies the evaluation of the Hartree-Fock mean-field energy of the system. We also assume that the system is trapped by a triaxial anisotropic harmonic potential given by 
\begin{equation}
 V_{\rm trap}({\bf r})=\frac{M}{2}\left(\omega_{x}^2{x}^2+\omega_{y}^2{y}^2+\omega_{z}^2{z}^2\right)\, ,
 \label{eq:trap}
\end{equation}
where $\omega_i$ with $i=x, y, z$ denote the respective trap frequencies.

\subsection{Wigner function in equilibrium}
\label{sec:wfeq}

The physical properties of the above described system can be captured by means of the semiclassical Wigner function \cite{goral}. Indeed, the quantum-mechanical expectation values of the system observables, which are required for the calculation of the properties of nonrelativistic quantum systems based on exact diagonalization, can be obtained as their phase-space averages, weighted by the Wigner function. In the case where the dipolar orientation axis lies along one of the trapping axes, an accurate ansatz for the Wigner function takes the simple form \cite{Sogo, HeZhang2Yi, Sogo2, Sogo2c, Baillie, Zhang,Falk, Veljic1,lima1,lima2},
\begin{equation}
 \Theta\left(1-\sum_{i} \frac{r_i^2} {R_i^2}-\sum_{i} \frac{k_i^2}{K_i^2}\right)\, ,
 \label{eq:ansatzwignerOld}
\end{equation}
where $\Theta$ represents the Heaviside step function, while the variational parameters $R_i$ and $K_i$ stand for the Thomas-Fermi radius and the Fermi momentum in the direction $i=x,y,z$, respectively. This ansatz is motivated by the Fermi-Dirac distribution of a noninteracting Fermi gas at zero temperature, whose Wigner function has the above form. A theory based on the above ansatz \cite{Falk, Sogo} was successfully used to determine the deformation of the FS, while its extension \cite{Veljic1} enabled a detailed analysis of the ground state and the time-of-flight (TOF) expansion dynamics of the system for different collisional regimes. Furthermore, numerical comparisons \cite{ZhangYi, RonenBohn} have confirmed that, even in the case of polar molecules with masses of the order of 100 atomic units and an electric dipole moment as large as 1~D, the above variational ansatz yields highly accurate results, within a fraction of per mille. This indicates that the general ansatz (\ref{eq:ansatzwignerOld}), first introduced in a slightly different manner in reference \cite{Sogo}, is indeed very well suited to describe dipolar Fermi gases.

However, the experiment of \cite{Francesca} was performed for an arbitrary angle $\theta$, and therefore the comparison of the theory \cite{Falk, Veljic1} was only possible for the special case of dipoles oriented along the $z$ axis, i.e., for $\theta=0^\circ$. Therefore, in order to model the global equilibrium distribution of the dipolar Fermi gas for arbitrarily oriented dipoles and to provide an accurate description of the experiment, it is necessary to generalise the ansatz (\ref{eq:ansatzwignerOld}). This is done in the present paper, where we apply an analogous reasoning and introduce the following ansatz for the Wigner distribution
\begin{equation}
 \nu^{}({\bf r},{\bf k})=\Theta\left(1-\sum_{i,j} r_i\mathbb{A}_{ij}r_j-\sum_{i,j} k_i\mathbb{B}_{ij}k_j\right)\, ,
\label{eq:ansatzwigner}
\end{equation}
where ${\mathbb{A}}_{ij}$ and $\mathbb{B}_{ij}$ are matrix elements that account for the generalised geometry of the system and determine the shape of the cloud in real space and of the FS in momentum space, respectively.

\begin{figure}[!b]
\centering
\includegraphics[width=14cm]{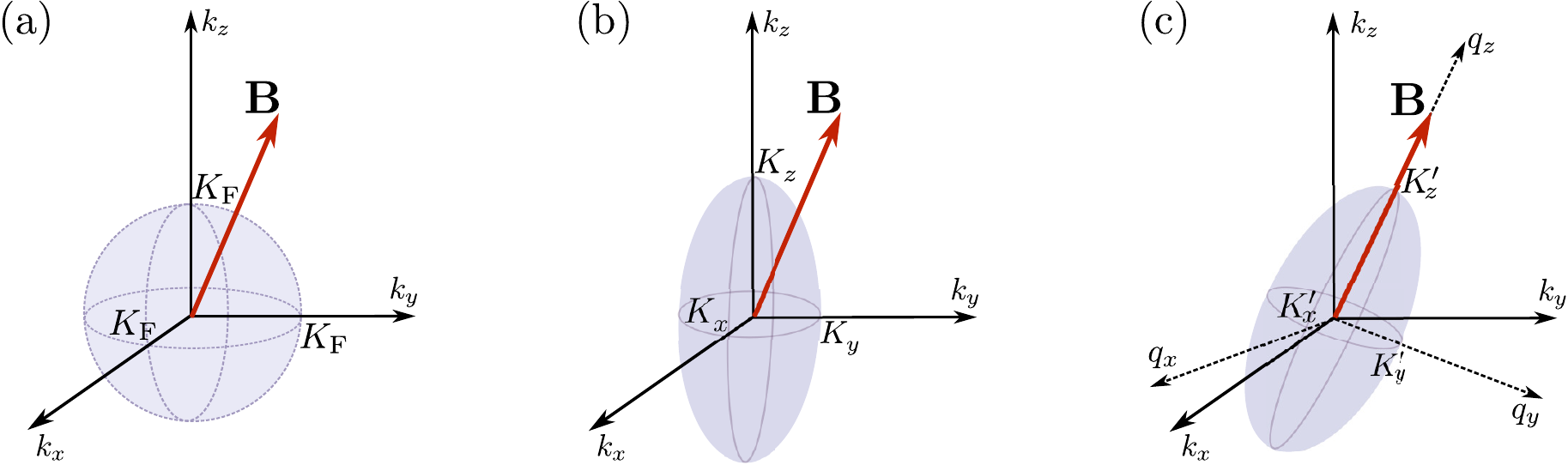}
\caption{Schematic illustration of possible FS configurations in the presence of an external field $\bf B$, described by spherical coordinates ${\bf B}=(B,\theta, \varphi)$. (a) Spherical scenario: FS remains spherical. (b) On-axis scenario: FS is deformed into an ellipsoid whose axes coincide with the trap axes. (c) Off-axis scenario: FS is deformed into an ellipsoid stretched in the direction of $\bf B$.}
\label{fig:schemas}
\end{figure} 

The particle density distribution in real space is determined by the trapping potential and the Hartree direct energy. Therefore, one expects that the matrix $\mathbb{A}$ can be well approximated by a diagonal matrix in the coordinate system $S$, which is defined by the harmonic trap axes,
\begin{equation}
\mathbb{A}=\left( \begin{array}{ccc}
1/R_x^2 & 0 & 0 \\
0 & 1/R_y^2 & 0 \\
0 & 0 & 1/R_z^2 \end{array} \right)\, .
\label{eq:mA}
\end{equation}
In this way, we neglect off-diagonal elements, which may arise due to the dipoles' arbitrary orientation. However, this is certainly justified for elongated traps, for which the cloud shape in the ground state is well determined by the trap. Therefore, we will consider here trap configurations that satisfy this condition.

On the other side, the momentum distribution is dominated by the interplay between the Pauli pressure, which is isotropic, and the Fock exchange energy, which is responsible for the deformation of the FS~\cite{Sogo}. The experiment of reference \cite{Francesca} suggested that the FS follows the rotation of the external field, keeping the major axis of the FS always parallel to the direction of the maximum attraction of the DDI. Motivated by this, we will consider several possible scenarios for a detailed theoretical description, in order to verify the above hypothesis.

For the sake of completeness, we start with a simple spherical scenario, in which the FS remains a sphere, as displayed in figure~\ref{fig:schemas}(a). In that case all Fermi momenta are equal ($K_i=K_F$) and the matrix $\mathbb{B}$ is given by $\mathbb{B}_1=\mathbb{I}/K_{F}^2$, where $\mathbb{I}$ is the $3\times 3$ identity matrix. We also consider a second, on-axis scenario, which includes the FS deformation such that it is an ellipsoid with fixed major axes coinciding with the trap axes, as shown in figure~\ref{fig:schemas}(b). Here the matrix $\mathbb{B}$ has a diagonal form,
\begin{equation}
\mathbb{B}_2=\left( \begin{array}{ccc}
1/K_x^{2} & 0 & 0 \\
0 & 1/K_y^2 & 0 \\
0 & 0 & 1/K_z^2 \end{array} \right)\, ,
\label{eq:mB}
\end{equation}
in a similar way as the matrix $\mathbb{A}$, which recovers the old ansatz given by (\ref{eq:ansatzwignerOld}). We note that the first, spherically-symmetric scenario is a special case of the second ansatz, obtained by restricting the Fermi momenta to be equal. Finally, as a third and more general possibility, we consider the off-axis hypothesis of \cite{Francesca} and assume that the matrix $\mathbb{B}$ has a diagonal form  $\mathbb{B}'_3$ in a rotated coordinate system $S'$, which is defined by the axes $q_x$, $q_y$ and $q_z$, where the $q_z$ axis remains parallel to the dipole moments, as depicted in figure~\ref{fig:schemas}(c),
\begin{equation}
\mathbb{B}'_3=\left( \begin{array}{ccc}
1/K_x'^{2} & 0 & 0 \\
0 & 1/K_y'^2 & 0 \\
0 & 0 & 1/K_z'^2 \end{array} \right)\, .
\label{eq:mBp}
\end{equation}
Here the parameters $K_i'$ represent the Fermi momenta in the rotated coordinate system $S'$. In order to describe the rotation from $S$ to $S'$, we introduce a rotation matrix $\mathbb{R}$,
\begin{equation}
\mathbb{R}=\left( \begin{array}{ccc}
\cos\theta\cos\varphi & -\sin\varphi & \sin\theta\cos\varphi \\
\cos\theta\sin\varphi & \cos\varphi & \sin\theta\sin\varphi \\
-\sin\theta & 0 & \cos\theta \end{array} \right)\, ,
\label{eq:mR}
\end{equation}
such that $\mathbb{B}'_3=\mathbb{R}^T\mathbb{B}_3\mathbb{R}$ and ${\bf q}=\mathbb{R}^T \bf k$, where the angles $\theta$ and $\varphi$ are defined in figure~\ref{fig:exp_setup}. We again note that the on-axis scenario is a special case of the off-axis one when the dipoles are parallel to one of the trap axes. For example, for $\theta=\varphi=0^\circ$ the matrix $\mathbb{B}_3$ is already diagonal, i.e., $K'_i=K_i$. We also note that, for all considered ans\"atze, the normalisation of the Wigner distribution $\nu({\bf r},{\bf k})$ is given by
\begin{equation}
\label{eq:partconser}
 N=\int\hspace*{-3mm}\int \frac{d^3r d^3 k}{(2\pi)^3}\nu({\bf r},{\bf k})=\frac{\bar{R}^3\bar{K}'^3}{48}\, ,
\end{equation}
where the bar denotes the geometric averaging: $\bar{R}=\left(R_xR_yR_z\right)^{1/3}$ and $\bar{K}'=\left(K'_xK'_yK'_z\right)^{1/3}$.

To determine the values of the variational parameters for each scenario, as usual, we require that they minimise the total Hartree-Fock energy of the system. 
This leads, together with the particle number conservation (\ref{eq:partconser}), to algebraically self-consistent equations that determine the Thomas-Fermi radii and momenta. In section \ref{sec:toten} we calculate the total energy of the system for each of the outlined scenarios and then in section \ref{sec:resmin} we proceed to determine the one that yields a minimal energy and that will be used in the rest of the paper. We note that one can certainly consider even more general ans\"atze, however, as we will see from the comparison with the experimental data, the proposed approach is fully suitable for describing our system not only qualitatively, but also quantitatively.

\subsection{Total energy}
\label{sec:toten}

Now that we have identified several relevant ans\"atze for describing the Wigner function of a dipolar Fermi gas of tilted dipoles, we proceed to determine the optimal values of the variational parameters. In order to do so, we have to minimise the total energy of the many-body Fermi system, which is in the Hartree-Fock mean-field theory given by the sum of the kinetic energy $E_{\rm kin}$, the trapping energy $E_{\rm trap}$, the Hartree direct energy $E_{\rm dd}^{\rm D}$, and the Fock exchange energy $E_{\rm dd}^{\rm E}$. Within a semiclassical theory, they can be written in terms of the Wigner function according to \cite{goral}
\begin{eqnarray}
 E_{\rm kin}=&\int\hspace*{-3mm}\int \frac{d^3rd^3k}{(2\pi)^3}\frac{\hbar^2 {\bf k}^2}{2M}\nu({\bf r},{\bf k})\, ,  \label{Ekinf0} \\ 
 E_{\rm trap}=&\int\hspace*{-3mm}\int \frac{d^3r d^3k}{(2\pi)^3} V_{\rm trap}({\bf r})\nu({\bf r},{\bf k})\, ,\label{Etrapf0}  \\
 E_{\rm dd}^{\rm D}=&\frac{1}{2}\int\hspace*{-3mm}\int \hspace*{-3mm}\int\hspace*{-3mm}\int \frac{d^3r d^3r' d^3k d^3 k'}{(2\pi)^6}V_{\rm dd}({\bf r}-{\bf r'}) \nu({\bf r},{\bf k})  \nu({\bf r'},{\bf k'})\, , \label{Edf0}  \\
 E_{\rm dd}^{\rm E}=&-\frac{1}{2}\int\hspace*{-3mm}\int \hspace*{-3mm}\int\hspace*{-3mm}\int\frac{d^3r d^3r' d^3k d^3 k'}{(2\pi)^6}  V_{\rm dd}({\bf r'}) {\rm e}^{i({\bf k}-{\bf k'}) \cdot {\bf r'}} \nu({\bf r},{\bf k}) \nu({\bf r},{\bf k'})\, , \label{Eexf0}
\end{eqnarray}
and have already been calculated before with an ansatz (\ref{eq:ansatzwignerOld}) for the case when the dipoles are parallel to one of the trap axes \cite{Sogo,lima1,lima2,Falk,Veljic1}. Whereas both the kinetic energy (\ref{Ekinf0}) and the trapping energy (\ref{Etrapf0}) yield simple integrals, the computation of the integrals in the Hartree term (\ref{Edf0}) and the Fock term (\ref{Eexf0}) is nontrivial and is therefore presented for the most general case in \ref{sec:HE} and \ref{sec:FE}, respectively.

In the spherical scenario, depicted in figure~\ref{fig:schemas}(a), the total energy of the system can be calculated using $K_i =K_{F}$ in ansatz (\ref{eq:ansatzwigner}), where the Fock exchange term turns out to give no contribution, yielding 
\begin{equation}
 E_{\rm tot}^{\rm (1)}=\frac{N}{8} \left(\frac{3\hbar^2K_{\rm F}^2}{2M}+\sum_j\frac{M \omega_j^2R_j^2}{2}\right) -\frac{6N^2 c_0}{ \bar{R}^3}f_{\rm A}\left( \frac{R_x}{R_z},\frac{R_y}{R_z},\theta,\varphi \right)\, .
 \label{Etot_b}
\end{equation}
Here $c_0=2^{10}C_{\rm dd}/(3^4\cdot 5\cdot 7\cdot \pi^3)$ is a constant related to the dipolar strength, while the features of the DDI are embodied into the generalised anisotropy function $f_{\rm A}(x,y,\theta,\phi)$, which includes explicitly the angular dependence of the DDI. It is defined as
\begin{equation}
\fl f_{\rm A}\left(x,y,\theta,\varphi\right)=\sin^2\theta\cos^2\varphi f\left( \frac{y}{x},\frac{1}{x} \right)  +\sin^2\theta\sin^2\varphi   f\left( \frac{x}{y},\frac{1}{y} \right)  +\cos^2\theta f\left( {x},{y} \right)\, ,
\label{gen_aniso_func}
\end{equation}
where $f(x,y)$ stands for the well-known anisotropy function derived, at first, for dipolar BECs \cite{pfau_aniso_func}. Note that $f(x,y)=f_A(x,y,0,0)$. This function has been encountered also in previous studies of fermionic dipolar systems \cite{lima2} in the hydrodynamic collisional regime, as well as in the transition from the collisionless to the hydrodynamic regime in both the TOF expansion dynamics \cite{Veljic1} and collective excitations \cite{Falk}. More details on the anisotropy and the generalised anisotropy function are given in \ref{sec:af}.

In the on-axis scenario, the FS is deformed to an ellipsoid whose axes are taken to be parallel to the trap axes, as shown in figure~\ref{fig:schemas}(b). This ansatz leads to the total energy of the system given by
\begin{eqnarray}
E_{\rm tot}^{\rm (2)}=&&\frac{N}{8}\sum_j \left(\frac{\hbar^2K_j^2}{2M}+\frac{M \omega_j^2R_j^2}{2}\right) \nonumber\\
&&-\frac{6N^2 c_0}{ \bar{R}^3} \left[f_{\rm A}\left( \frac{R_x}{R_z},\frac{R_y}{R_z},\theta,\varphi \right)-f_{\rm A}\left( \frac{K_z}{K_x},\frac{K_z}{K_y},\theta,\varphi \right)\right]\, .
 \label{Etot_c}
\end{eqnarray}
Note that (\ref{Etot_c}) reduces, indeed, to (\ref{Etot_b}) for the special case of $K_x=K_y=K_z$, since $f_{\rm A}(1, 1,\theta,\varphi)=0$.

Finally, in the most-general considered off-axis scenario displayed in figure~\ref{fig:schemas}(c), we allow for both the FS deformation and its rotation so that one of its axes is parallel to the external field. In this case, the total energy of the system reads as
\begin{eqnarray}
E_{\rm tot}^{\rm (3)}=&&\frac{N}{8}\sum_j \left(\frac{\hbar^2K_j'^2}{2M}+\frac{M \omega_j^2R_j^2}{2}\right) \nonumber \\
&&-\frac{6N^2 c_0}{ \bar{R}^3} \left[f_{\rm A}\left(\frac{R_x}{R_z},\frac{R_y}{R_z},\theta,\varphi\right) -f\left( \frac{K_z'}{K_x'},\frac{K_z'}{K_y'}\right) \right]\, . \label{Etot_d}
\end{eqnarray}
Note that the above form of the Fock energy in the last term is not surprising if we bear in mind the form of equation (\ref{gen_aniso_func}). Namely, in the rotated coordinate system $S'$, the axis $q_z$ coincides with the direction of the external field, so the generalised anisotropy function $f_A$ reduces to the standard anisotropy function $f$, just with the arguments $K'_i$ from the rotated system.

\section{Ground-state properties}
\label{sec:res}

Having obtained the total energy for all three scenarios, we now determine which configuration minimises the system's total energy for a fixed particle number and, hence, can be considered as the most physically suitable ansatz for the ground state of the system of dipolar fermions. Afterwards, we use it to numerically calculate the FS and atomic cloud deformation, and discuss the obtained results.

In practical terms, we start from (\ref{Etot_b}), (\ref{Etot_c}) and (\ref{Etot_d}), and minimise the energy of the system for each scenario under the constraint (\ref{eq:partconser}) that the particle number $N$ is fixed to a given value. Therefore, the corresponding equations are obtained by extremizing the grand-canonical potential $\Omega^{(k)}=E^{(k)}_\mathrm{tot}-\mu N$ for $k=1,2,3$ with respect to the variational parameters, where $\mu$ is the chemical potential of the system, and the particle number $N$ in the last term is replaced by the expression (\ref{eq:partconser}) when $\Omega^{(k)}$ is evaluated. In this way, the chemical potential acts as a Lagrange multiplier and fixes the particle number through the condition $N=-\partial\Omega^{(k)}/\partial\mu$.

In the spherical scenario, there are five variational parameters, ($K_F$, $R_i$, $\mu$), where $i=x,y,z$. The corresponding five equations are obtained by setting the first derivatives of $\Omega^{(1)}$ with respect to 
$K_F$ and $R_i$ to zero, plus the particle-number fixing equation, i.e., $N=-\partial\Omega^{(1)}/\partial\mu$. In both the on-axis and the off-axis scenario we have seven variational parameters: ($K_i$, $R_i$, $\mu$) and ($K'_i$, $R_i$, $\mu$), respectively. The sets of seven equations for both cases are obtained similarly as in the previous case. The complete sets of equations for the respective variational parameters for all cases are given in \ref{var_param}.

\subsection{Minimisation of total energy}
\label{sec:resmin}

\begin{figure}[!b]
\centering
\includegraphics[width=15cm]{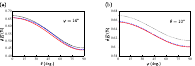}
\caption{Relative total energy shift (\ref{eq:deltaE}) for experimental parameters of \cite{Francesca} as a function of: (a) angle $\theta$, for fixed $\varphi=14^\circ$; (b) angle $\varphi$ for fixed $\theta=10^\circ$. The three curves, correspond to $\delta E^{\rm (1)}$ (black dotted line, spherical scenario), $\delta E^{\rm (2)}$ (blue dashed line, on-axis scenario), and $\delta E^{\rm (3)}$  (red solid line, off-axis scenario).}
\label{fig:del_E}
\end{figure}

In order to compare the three ans\"atze, we solve the corresponding sets of equations and calculate the total energy of the system in each case. As a model system, we consider the case of a dipolar Fermi gas of atomic $^{167}$Er, using the typical values from the Innsbruck experiments (see below and also \cite{Francesca}), $N=6.6\times 10^4$, ($\omega_x$, $\omega_y$, $\omega_z$) = (579, 91, 611) $\times 2\pi$~Hz, unless otherwise specified.
The underlying geometry of the experimental setup is depicted in figure~\ref{fig:exp_setup}.

In figure~\ref{fig:del_E} we compare the total energy of the system as a function of angles $\theta$ and $\varphi$ for all three different scenarios. The comparison is done in terms of the relative total energy shift
\begin{equation}
\delta E=\frac{E_{\rm tot}}{E_0}-1\, ,
\label{eq:deltaE}
\end{equation}
where $E_0=\frac{3}{4}NE_F$ stands for the total energy of the ideal Fermi gas confined into a harmonic trap (\ref{eq:trap}), and $E_F=\hbar \bar \omega (6N)^{1/3}$ denotes its Fermi energy, where $\bar{\omega}=(\omega_x \omega_y \omega_z)^{1/3}$. Figure~\ref{fig:del_E}(a) presents the relative total energy shifts as functions of the angle $\theta$ for a fixed value of the angle $\varphi=14^\circ$, corresponding to the typical experimental configuration (see below). The three curves, from top to bottom, correspond to $\delta E^{\rm (1)}$, $\delta E^{\rm (2)}$, and $\delta E^{\rm (3)}$, respectively. As a cross-check, we note that the total energies $E_{\rm tot}^{\rm (2)}$ and $E_{\rm tot}^{\rm (3)}$ coincide for $\theta=0^\circ$. This is expected, since the on-axis scenario is a special case of the off-axis one for $\theta=0^\circ$. From this figure we immediately see that there are no intersections between the curves, and that it always holds $E_{\rm tot}^{\rm (1)}\geq E_{\rm tot}^{\rm (2)}\geq E_{\rm tot}^{\rm (3)}$. As a consequence, we conclude that the off-axis scenario, in which the FS is deformed into an ellipsoid that follows the orientation of the dipoles, is favoured among the considered cases as it has the minimal energy. The same conclusion is obtained if we consider the $\varphi$-dependence of the relative total energy shifts, depicted in figure~\ref{fig:del_E}(b) for a fixed value of the angle $\theta=10^\circ$. More exhaustive numerical calculations show that this remains to be true even for arbitrary values of the angles $\theta$ and $\varphi$.

Comparing figures~\ref{fig:del_E}(a) and \ref{fig:del_E}(b) we see that the relative total energy shift always remains small, of the order of $0.4-0.7\%$, due to a relatively weak DDI between the erbium atoms compared to the energy scale set by the Fermi energy. We also see that the $\theta$-dependence of the total energy is much stronger than the corresponding $\varphi$-dependence. We note that the shift would certainly be more significant for atomic and molecular species with a stronger DDI.

The above conclusion is valid not only for the parameters used in figure~\ref{fig:del_E}, but, in fact, we have numerically verified that the off-axis scenario for the ansatz (\ref{eq:ansatzwigner}) for the Wigner function in global equilibrium always yields a minimal energy given by (\ref{Etot_d}) and, thus, we will use it throughout the rest of the paper. The corresponding equations determining all seven variational parameters are given in \ref{var_param} as (\ref{VSTKx})--(\ref{VSTN}). A closer examination of those equations reveals that the FS is always a cylindrically symmetric ellipsoid, where $K_x'=K_y'$ holds. This is expected, since the orientation of the dipoles in the rotated coordinate system coincides with the $q_z$ axis and singles this particular direction out, leaving the perpendicular plane perfectly symmetric in momentum space. Therefore, as shown in \ref{var_param}, the equations for the seven variational parameters can be rewritten in a more convenient form as (\ref{VSTRx})--(\ref{mu3}).

\subsection{Fermi surface deformation}

Now that we have shown that the FS is, indeed, deformed by the DDI into an ellipsoid, which follows the orientation of the dipoles, we study the angular dependence of this deformation in more detail. To that end, and taking into account the cylindrical symmetry of the FS, we define the FS deformation as the difference between the momentum-space aspect ratio for the dipolar and the noninteracting Fermi gas in the rotated system $S'$ according to
\begin{equation}
\Delta=\frac{K_z'}{K_x'}-1\, .
\label{delta}
\end{equation}
This quantity measures the degree of deformation, which emerges purely due to the DDI. In particular, we investigate how the deformation $\Delta$ depends on the trap geometry, the orientation of the dipoles, the DDI strength, and the number of particles. The DDI strength is usually expressed in terms of a dimensionless relative strength $\varepsilon_{\rm dd}$, defined as
\begin{equation}
 \varepsilon_{\rm dd}=\frac{C_{\rm dd}}{4 \pi}\sqrt{\frac{M^3\bar{\omega}}{\hbar^5}}N^{1/6}\, ,
\end{equation}
which gives a rough estimate of the ratio between the mean dipolar interaction energy and the Fermi energy. We will use it to characterise the strength of the DDI when comparing its effects for different species.

\begin{figure}[!t]
\centering
\includegraphics[width=14.5cm]{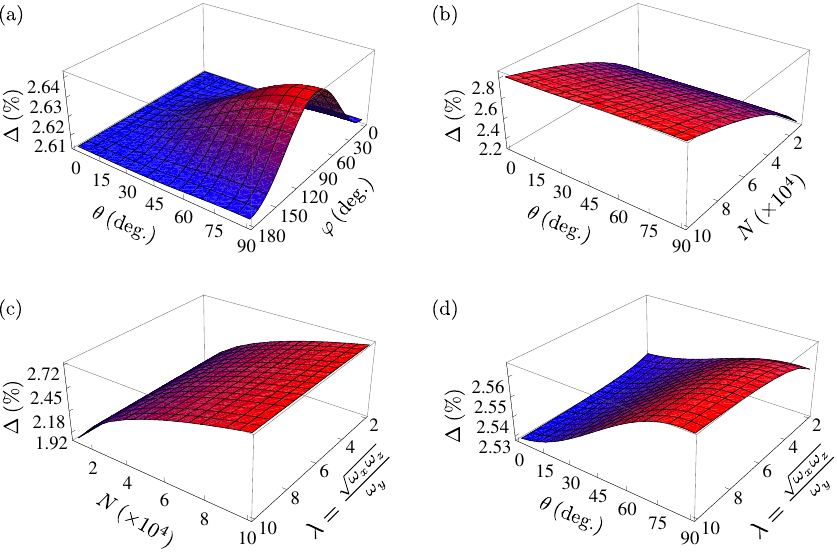}
\caption{FS deformation $\Delta$ for $^{167}$Er atoms with magnetic dipole moment $m=7\mu_{\rm B}$ as a function of: (a) dipoles' orientation angles $\theta$ and $\varphi$, for parameters of \cite{Francesca}; (b) particle number $N$ and angle $\theta$, for $\varphi=90^\circ$ and trap frequencies of \cite{Francesca}; (c) particle number $N$ and trap anisotropy $\lambda$, for $\theta=\varphi=90^\circ$; (d) angle $\theta$ and trap anisotropy $\lambda$, for $\varphi=90^\circ$ and $N=6.6\times 10^4$. Trap anisotropy $\lambda$ in (c) and (d) was varied by changing the frequencies $\omega_x=\omega_z$ and $\omega_y$, while keeping the mean frequency $\bar{\omega}=300\times 2\pi$~Hz constant.}
\label{fig:FS_Er}
\end{figure}

We now calculate the FS deformation of $^{167}$Er for the parameters of \cite{Francesca}, yielding the relative interaction strength $\varepsilon_{\rm dd}=0.15$.
In figure~\ref{fig:FS_Er}(a) we present the angular dependence of $\Delta$ on $\theta$ and $\varphi$, whose values turn out to be around $2.6\%$, consistent with earlier experimental results \cite{Francesca}. We observe that there is a maximum deformation of the FS at $\theta=\varphi=90^\circ$, which  corresponds to the direction of the smallest trapping frequency $\omega_y$ ($y$ axis). This can be understood heuristically, if one recalls that the DDI is attractive for dipoles oriented head-to-tail. Thus, a weaker trapping frequency favours the stretching of the gas in that direction so that, in turn, this cigar-shaped configuration enhances the relative contribution of the DDI to the total energy.

Another aspect relevant for experiments is the influence of the particle number $N$ and the trap geometry on the deformation of the FS. Tuning these parameters and the direction of the dipoles might lead to an enhancement of the DDI effects, and therefore to a stronger deformation of the FS. This is investigated in figures~\ref{fig:FS_Er}(b)-\ref{fig:FS_Er}(d), where the FS deformation is given as a function of parameters $N$, $\theta$ and the trap anisotropy $\lambda=\sqrt{\omega_x \omega_z}/\omega_y$ for a fixed value of the angle $\varphi=90^\circ$. Figures~\ref{fig:FS_Er}(c) and \ref{fig:FS_Er}(d) explore the FS deformation as a function of the trap anisotropy $\lambda$, which was varied by changing the frequencies $\omega_x=\omega_z$ and $\omega_y$, while keeping the mean frequency $\bar{\omega}=300\times 2\pi$~Hz constant. From all these figures we conclude that the increase in the particle number yields a dominant increase in $\Delta$ compared to all other parameters. We note that, in fact, $\Delta$ also depends on $\bar{\omega}$, which we do not show here, since it can be directly connected to the particle number dependence. Indeed, the FS deformation depends on $\varepsilon_{\rm dd}$ \cite{Francesca}, yielding a dependence of $\Delta$ on $N^{1/6}\bar{\omega}^{1/2}$. As the trap frequencies can be more easily tuned than the particle number, $\bar{\omega}$ can be considered as a predominant control knob in the experiment. However, a precise control of the angles and the anisotropy, which is experimentally easy to realise, may help to achieve an even larger increase in the deformation of the FS. We also note that the $\lambda$-dependence is the weakest one, and therefore the formalism for calculating the angular dependence presented here is significant for a systematic study of the influence of the relevant parameters. 

\begin{figure}[!b]
\centering
\includegraphics[width=7cm]{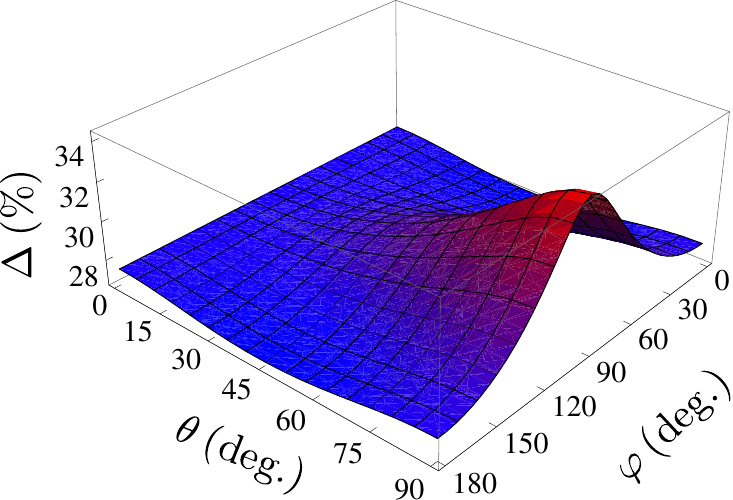}
\caption{Angular dependence of FS deformation $\Delta$ for polar molecules with electric dipole moment $m = 0.25\:\rm D$ and relative DDI strength $\varepsilon_{\rm dd}=1.51$ for parameters of \cite{Francesca}.}
\label{fig:FS_KRb}
\end{figure}

Finally, we study the role of the DDI strength and explore whether qualitative changes of the system's behaviour emerge by increasing the value of the dipole moment. To this aim, we compare the erbium case with a molecular Fermi gas of $^{40}$K$^{87}$Rb, assuming that the same gas characteristics can be achieved in the same trap with this species \cite{DSJin}. The latter possesses an electric dipole moment of strength $m=0.56\:\rm D$, yielding a much larger relative interaction strength $\varepsilon_{\rm dd}=7.76$ for the same parameters. Since the critical value of $\varepsilon_{\rm dd}$, for which the system is stable, amounts to $\varepsilon_{\rm dd}^{\rm crit}=2.5$ \cite{Veljic1}, the molecular $^{40}$K$^{87}$Rb gas in such a geometry and with the maximal strength of the DDI would in fact not be stable and would collapse under the attractive action of the DDI. For the sake of simplicity and comparison between the systems, we consider a molecular sample of similar geometry and atom number but in which the dipole moment has been tuned to $m = 0.25\:\rm D$ by means of an external field \cite{sigma}. This leads to the relative DDI strength $\varepsilon_{\rm dd}=1.51$, which we use in the following.

As we see, the FS deformation $\Delta$ has a much stronger angular dependence in figure~\ref{fig:FS_KRb} than in figure~\ref{fig:FS_Er}(a) for the erbium case. Furthermore, in figure~\ref{fig:FS_KRb} we see that $\Delta$ has a local minimum for $\varphi=0^\circ$ around $\theta=40^\circ$, while no such minimum exists in figure~\ref{fig:FS_Er}(a). Only a detailed numerical study based on the formalism developed here can provide a precise landscape of the FS deformation behaviour for a concrete experimental setup.

Although the shapes of both angular dependencies in figures~\ref{fig:FS_Er}(a) and \ref{fig:FS_KRb} are quite similar, the main difference is that the deformation of the FS for polar molecules is an order of magnitude larger than for erbium and has a value of around $30\%$. However, we also observe that the variation in the values of $\Delta$ for different angles $\theta$ and $\varphi$ is around $0.03\%$ in the case of an atomic erbium gas, while for the molecules it amounts to around $5\%$, i.e., the variations of $\Delta$ are two orders of magnitude larger for the molecular case. The reason for this increase in both the maximal FS deformation and its angular variation is the same, namely the increase in the relative DDI strength $\varepsilon_{\rm dd}$, which is one order of magnitude larger for our molecules compared to $^{167}$Er. While the FS deformation is proportional to $\varepsilon_{\rm dd}$, as expected \cite{Veljic1} and as evidenced by our results above, our findings suggest that its maximal angular variation is proportional to $\varepsilon_{\rm dd}^2$.

\begin{figure}[!b]
\centering
\includegraphics[width=12cm]{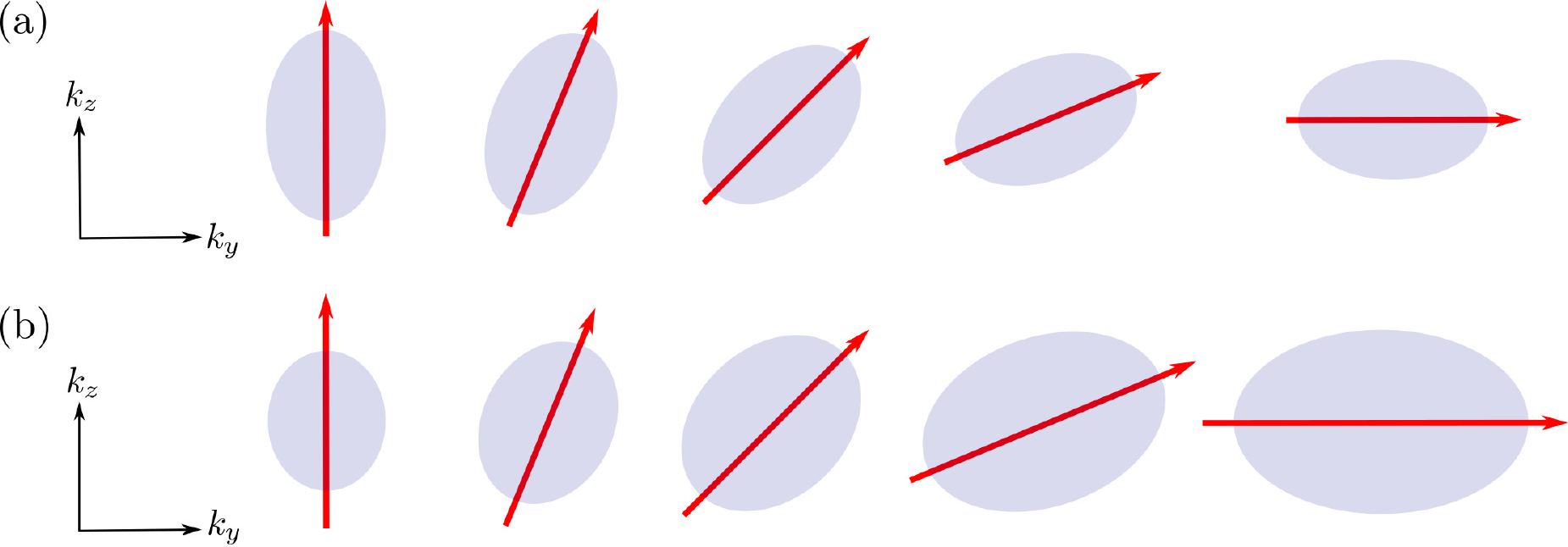}
\caption{Illustration of the angular dependence of the FS deformation in momentum space for system in an anisotropic trap elongated along the horizontal $y$ axis (see figure~\ref{fig:exp_setup}): (a) for weak DDI, when the FS ellipsoid just rotates like a rigid object; (b) for strong DDI, when the FS ellipsoid stretches in all directions and its deformation strongly depends on the dipoles' orientation.}
\label{fig:FS_soft_rigid}
\end{figure}

The calculated angular dependence of the FS deformation on the DDI strength has the following important physical consequence. For erbium atoms, where $\varepsilon_{\rm dd}$ is small, the angular variation of the FS deformation is even smaller, since it is proportional to $\varepsilon_{\rm dd}^2$, and it would be difficult to observe in experiments. Therefore, one could say that the FS behaves as a rigid ellipsoid, which just rotates following the orientation of the dipoles, without changing its shape \cite{Francesca}, as illustrated in figure~\ref{fig:FS_soft_rigid}(a). This also implies that the atomic cloud shape in real space is practically disentangled from the FS, and is mainly determined by the trap shape. On the other hand, when $\varepsilon_{\rm dd}$ is large enough, as in the case of $^{40}$K$^{87}$Rb, the FS not only rotates, but also significantly changes its shape, since the angular variation can be as high as $5\%$, which is experimentally observable. This is schematically shown in figure~\ref{fig:FS_soft_rigid}(b), where the FS behaves as a soft ellipsoid, whose axes are stretched as it rotates. Although we know that the phase-space volume is preserved, according to the particle number conservation (\ref{eq:partconser}), figure~\ref{fig:FS_soft_rigid}(b) illustrates that the FS, i.e., the momentum-space volume increases ($K_i'$ increase), while in real space the volume of the cloud shape decreases ($R_i$ decrease). From this we see that the real-space atomic cloud shape is indeed coupled to the FS, and this effect can become measurable in future dipolar fermion experiments, with sufficiently large values of $\varepsilon_{\rm dd}$.

\subsection{Real-space magnetostriction}

The presence of the DDI in both bosonic \cite{yi_s_2001} and fermionic \cite{goral} quantum gases has been predicted and evidenced in experiment by detailed TOF expansion measurements \cite{Cr-magnetostriction} to induce magnetostriction in real space, i.e., a stretching of the gas cloud along the direction of the dipoles. In this section we investigate the dependence of this effect on the orientation of the dipoles for the fermionic case. To this end, we first define real-space aspect ratios $A_{ij}=R_i/R_j$ of the corresponding Thomas-Fermi radii, as well as their noninteracting counterparts $A^0_{ij}=R^0_i/R^0_j=\omega_j/\omega_i$. The magnetostriction can now be studied in terms of the relative cloud deformations:
\begin{equation}
\delta_{xz}=A_{xz}/A^0_{xz}-1\, , \quad \delta_{yz}=A_{yz}/A^0_{yz}-1\, .
\label{delta_ar_real_space}
\end{equation}
Here the anisotropies due to the harmonic trap are already taken into account and eliminated, so that only effects of the DDI contribute to the nontrivial value of $\delta_{xz}$ and $\delta_{yz}$. This is in close analogy to the definition of the relative total energy shift of the system in (\ref{eq:deltaE}), or the FS deformation in (\ref{delta}).

\begin{figure}[!b]
\centering
\includegraphics[width=7cm]{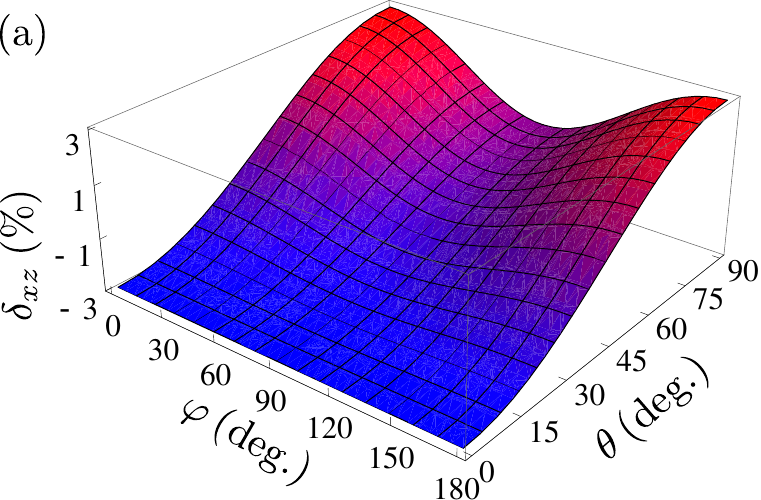} \hspace{4mm}
\includegraphics[width=7cm]{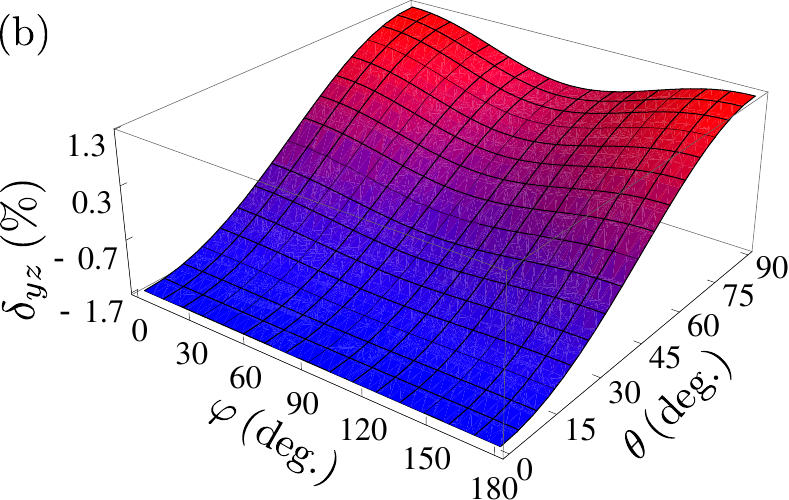}
\caption{Angular dependence of relative cloud deformation (\ref{delta_ar_real_space}) for $^{167}$Er, with parameters as in figure~\ref{fig:FS_Er}(a): (a) $\delta_{xz}$; (b) $\delta_{yz}$.}
\label{fig:asp_rat_real_sp}
\end{figure}

In figure~\ref{fig:asp_rat_real_sp} we present the angular dependence of the relative cloud deformations for the Fermi gas of erbium with the same parameters as in figure~\ref{fig:FS_Er}(a). We see that both deformations for a fixed angle $\theta>0$ turn out to possess a minimum for $\varphi=90^\circ$, while along the $\varphi$ direction the FS deformation monotonously increases. If we compare this to the behaviour in momentum space, see figure~\ref{fig:FS_Er}(b), we see that along the $\varphi$ direction we have qualitatively the same type of increasing dependence, while the behaviour in the $\theta$ direction is markedly different. Indeed, in momentum space it exhibits a maximum for $\varphi=90^\circ$, while in real space we observe a minimum value in that direction for $\varphi=90^\circ$. A related effect has been previously found, showing that the Bose gas momentum becomes distorted in the opposite sense to that of the Fermi gas \cite{Baillie}. There, the effect can be traced back to the differences in the statistical nature of bosons and fermions. Here, however, the different behaviour is due to the anisotropic nature of the DDI and its interplay with the direction of the dipoles and the trap geometry.

\section{Comparison with the experiment}
\label{compa_exp}

Having developed a general theoretical framework and having numerically studied the ground-state properties of dipolar Fermi gases with arbitrary oriented dipoles in the two previous sections, we now compare those results with experimental data obtained in our experimental setup in Innsbruck, producing degenerate Fermi gases of erbium \cite{Er}. The FS deformation was first observed with this setup and reported in reference \cite{Francesca}. Here we perform additional measurements, using different trapping configurations to test our theoretical understanding developed above. In these experiments the FS deformation is probed by the TOF expansion measurements. Starting from a degenerate Fermi gas with $N\sim 6-7\times 10^4$ atoms and at the temperature $T/T_F\sim 0.2$, we slowly set the cloud geometry and dipole orientation to the one of interest, let the cloud equilibrate for several hundreds of milliseconds and then suddenly remove the trapping potential. After a free expansion of duration $t$, we perform standard absorption imaging along a fixed direction, see figure~\ref{fig:exp_setup}.

Before we compare theory and experiment, let us note again that our theoretical results are only valid for $T=0$. For finite temperatures the isotropic thermal fluctuations have already been shown to work against any directional dependence stemming from either the harmonic confinement or the DDI, thus they diminish the FS deformation. The thermal corrections to the total energy are known to be proportional to $(T/T_F)^2$ at low $T$ \cite{thermalLima}. The corresponding effect on the FS deformation was also previously theoretically \cite{thermalBlakie} and experimentally \cite{Francesca} investigated, showing similar scalings. However, for the low temperatures of our experiments, this would yield only a few percent correction to the zero-temperature results, which lies within the experimental error bars. Therefore, we can neglect the thermal corrections here. Generally speaking, the value of $(T/T_F)^2$ can be used to estimate the relevance of the finite-temperature effects for $T/T_F<0.5$, while $(T/T_F)^{-5/2}$ should be considered for larger temperatures \cite{thermalBlakie}.

\subsection{Aspect ratios and FS deformation}
\label{sec:ARandFS}

The TOF images are taken in the plane perpendicular to the imaging axis and the deformation of the atomic cloud is described in terms of the time-dependent cloud aspect ratio $A_{\rm R}$, which is defined as a ratio of vertical and horizontal radii of the cloud in the imaging plane. As depicted in figure~\ref{fig:exp_setup}, the imaging axis $y'$ is in the $xy$ plane, rotated by an angle $\alpha$ to the $y$ axis, and the aspect ratio $A_{\rm R}$ is given by \cite{Veljic1}
\begin{equation}
A_{\rm R}(t)=\sqrt{\frac{\left<r_z^2(t)\right>}{\left<r_x^2(t)\right>\cos^2\alpha+\left<r_y^2(t)\right>\sin^2\alpha}}\, ,
\label{eq:AR}
\end{equation}
where $\left<r_i^2(t)\right>$ is the average size of the cloud in the direction $i=x,y,z$ after TOF. These quantities are directly measurable in the experiment, and we use them to extract the value of the deformation of the FS, which is connected to the aspect ratio in momentum space. It is defined similarly as $A_{\rm R}$ \cite{Veljic1}, according to
\begin{equation}
A_{\rm K}=\sqrt{\frac{\left<k_z^2\right>}{\left<k_x^2\right>\cos^2\alpha+\left<k_y^2\right>\sin^2\alpha}}\, ,
\label{eq:AK0bal}
\end{equation}
where $\left<k_i^2\right>$ is the average size of the FS in the direction $i=x,y,z$ in global equilibrium, before the trap is released. Using the definition (\ref{eq:AK0bal}), a straightforward but lengthy calculation yields the following expression for the aspect ratio in momentum space in terms of the Fermi momenta $K'_i$ in the rotated coordinate system:
\small
\begin{equation}
\fl
A_{\rm K}=\sqrt{\frac{K_x'^2\sin^2\theta+K_z'^2\cos^2\theta}{K_x'^2[1-\sin^2\theta(\cos^2\varphi \cos^2\alpha +\sin^2\varphi \sin^2\alpha)]+K_z'^2\sin^2\theta(\cos^2\varphi \cos^2\alpha +\sin^2\varphi \sin^2\alpha)}}\, .
\label{eq:Ak}
\end{equation}
\normalsize
Please note that only for $\theta=0^\circ$, when the dipoles are parallel to the $z$ axis, the above momentum-space aspect ratio coincides with the ratio between the Fermi momenta, $A_{\rm K}=K_z'/K_x'=1+\Delta$, where $\Delta$ denotes the previously introduced deformation of the FS. In general, however, the relation between $A_{\rm K}$ and $\Delta$ is nonlinear and can be obtained from (\ref{eq:Ak}), as follows:
\begin{equation}
\Delta=\sqrt{\frac{A_{\rm K}^2[1-\sin^2\theta(\cos^2\varphi \cos^2\alpha + \sin^2\varphi\sin^2\alpha)]-\sin^2\theta}{\cos^2\theta-A_{\rm K}^2\sin^2\theta(\cos^2\varphi \cos^2\alpha +\sin^2\varphi \sin^2\alpha)}}-1\, .
\label{eq:deltaAk}
\end{equation}

In order to extract the value of the deformation of the FS from the experimental data using the above equation, we still need to calculate the momentum-space aspect ratio $A_{\rm K}$. This is done by using the fact that the long-time expansion is mainly dominated by the velocity distribution right after the release from the trap. Here we rely on the ballistic approximation, which assumes that the TOF images, that show the shape of the atomic cloud in real space, purely reflect the momentum distribution in the global equilibrium, i.e.,
\begin{equation}
\label{eq:AK0}
A_\mathrm{K}\approx\lim_{t\to\infty}A^\mathrm{bal}_\mathrm{R}(t)\, .
\end{equation}
In principle, this is true just in the case of ballistic expansion, when the effects of the DDI can safely be neglected during the TOF. However, since the DDI is long-range, it should be taken into account, rendering the TOF results always non-ballistic. A general theory that would allow such a treatment is not yet available and is beyond the scope of the present ground-state study. Nevertheless, if the DDI is weak enough, as in the case of erbium atomic gases, the difference between the ballistic (free) and non-ballistic expansion is small, as already shown in \cite{Veljic1}. Thus, (\ref{eq:AK0}) can approximately be used in our case and the value of $A_{\rm K}$ in global equilibrium can be extracted from the long-time limit of $A_{\rm R}$, which is available from the experimental data. We highlight that in some limiting cases it is still possible to take into account a non-ballistic expansion by using the previously developed dynamical theory \cite{Veljic1}. This is expected to yield a more precise value of the aspect ratio, as will be illustrated in the next section.

With those cautionary remarks in mind, we complete the algorithm for analysing our experimental data by calculating the FS deformation from the extracted aspect ratio using (\ref{eq:deltaAk}), which enables its comparison with our numerical results.

\subsection{Experimental and theoretical results}
\label{sec:exp_theory}

Here we consider three different datasets corresponding to the experimental parameters listed in table~\ref{tab:tab1}. Case 1 corresponds to the experimental results published in \cite{Francesca}, while Case 2 and Case 3 are new results reported here. Table~\ref{tab:tab1} also gives the mean frequency of the trap $\bar{\omega}$ and the trap anisotropy $\lambda$ for each case. While Cases 1 and 2 represent cigar-shaped traps, Case 2 is selected so that it has the same value of $\omega_y$ as Case 1, but a smaller anisotropy $\lambda$. On the other hand, Case 3 is chosen so that its mean frequency $\bar{\omega}$ is approximately the same as for Case 1, but its anisotropy $\lambda$ is much reduced. For each dataset, we probe the FS deformation for various angles $\theta$ and a fixed angle $\varphi=14^\circ$. The measurement for each experimental configuration is repeated a large number of times, typically twenty, so that the mean value can be reliably estimated and the statistical error is reduced below 0.2\%.

\begin{table}[!h]
\caption{\label{tab:tab1}
Number of atoms $N$, trap frequencies $\omega_i$, mean frequencies $\bar{\omega}$ and anisotropies $\lambda$ for three sets of experimental parameters used throughout this paper. Case 1 corresponds to \cite{Francesca}, while Case 2 and Case 3 are new results.}
\begin{indented}
\item[]
\begin{tabular}{lcllllc}
\br
$^{167}$Er& $N\: (\times 10^4)$& $\omega_x \:\rm (Hz)$ & $\omega_y \:\rm (Hz)$& $\omega_z \:\rm (Hz)$ & $\bar{\omega} \:\rm (Hz)$ & $\lambda$\\
\mr
Case 1 & 6.6 & 579$\times 2\pi$ & 91$\times 2\pi$ & 611$\times 2\pi$ & 318$\times 2\pi$ & 6.54\\
Case 2 & 6.3 & 428$\times 2\pi$ & 91$\times 2\pi$ & 459$\times 2\pi$ & 261$\times 2\pi$ & 4.87\\
Case 3 & 6.1 & 408$\times 2\pi$ & 212$\times 2\pi$ & 349$\times 2\pi$ & 311$\times 2\pi$ & 1.78\\
\br
\end{tabular}
\end{indented}
\end{table}

\begin{figure}[!t]
\centering
\includegraphics[width=7cm]{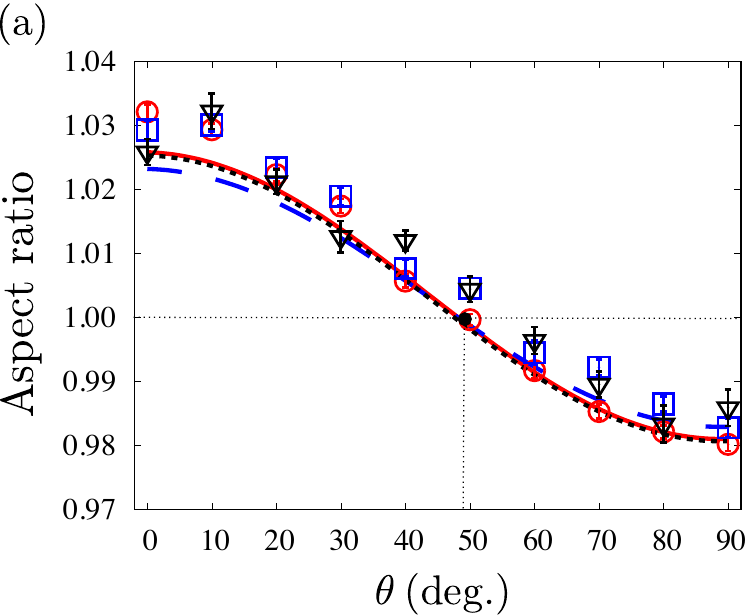} \hspace*{5mm}
\includegraphics[width=7cm]{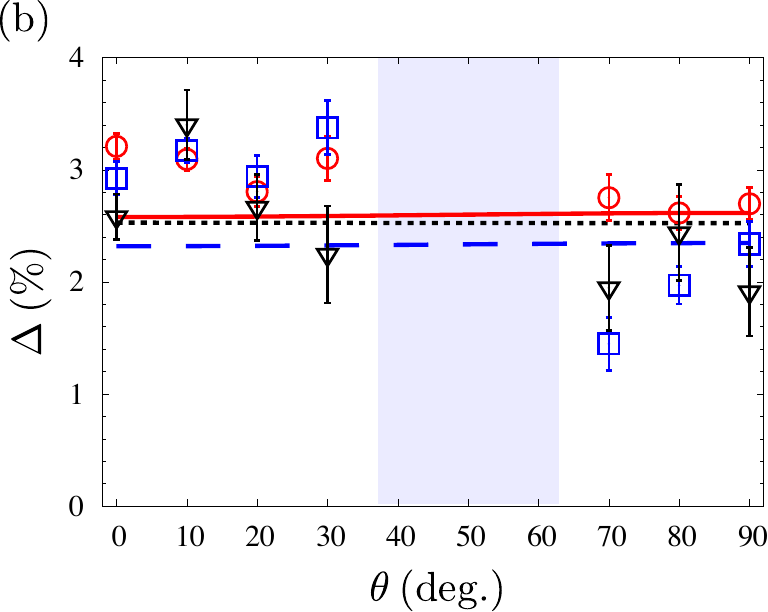}
\caption{Comparison of our results for $\theta$ dependence of: (a) theoretical value of aspect ratio $A_\mathrm{K}$ and its experimental estimate $A_{\rm R}^\mathrm{exp} (t=12\:\rm ms)$ according to (\ref{eq:AK0}); (b) theoretical value of the FS deformation $\Delta$ and its experimental estimate (see main text). Red solid lines and circles correspond to Case 1, blue dashed lines and squares correspond to Case 2, and black dotted lines and triangles correspond to Case 3. Vertical bars for experimental results correspond to statistical errors. Angle $\varphi=14^\circ$ was kept constant during the experiment. Intersection point of three curves in panel (a) corresponds to ($\theta^*$, $A_\mathrm{K}^*$)=($49.16^\circ$, 1), while shaded area in panel (b) is excluded due to a pole in (\ref{eq:err}); see main text for further details.}
\label{fig:comparison}
\end{figure} 

Figure~\ref{fig:comparison} shows a direct comparison between our experimental and theoretical results without any free parameters. Experimentally we measure the mean value of the aspect ratio $A_\mathrm{R}$ in free expansion using the TOF $t=12\:\rm ms$, which is taken to be sufficiently long so that the approximation (\ref{eq:AK0}) can be used, and yet not too long so that the cloud does not get too dilute and a reliable fit of the density distribution from the absorption images is possible. In figure~\ref{fig:comparison}(a) we show the $\theta$ dependence of the measured quantity $A_\mathrm{R}(12\,\mathrm{ms})$ for the parameters of Case 1 (red circles), Case 2 (blue squares) and Case 3 (black triangles), as well as the corresponding theoretical curves (solid red, dashed blue and dotted black line, respectively) for $A_\mathrm{K}$ at global equilibrium, calculated according to (\ref{eq:Ak}). We see that the agreement is generally very good, and that the experimental data are closely matched by the shape predicted by theory. At the same time, this figure also presents an a posteriori justification for using the ballistic approximation in those three cases.

\begin{table}[!t]
\caption{\label{tab:tab2}
Comparison of theoretical values of aspect ratios in momentum space $A_{\rm K}$ in global equilibrium and TOF aspect ratios in real space: theoretical value of $A_{\rm R}^{\rm nbal}$ and experimental value of $A_{\rm R}^{\rm exp}$, with corresponding statistical errors $\Delta A_{\rm R}^{\rm exp}$. Real-space aspect ratios correspond to TOF of $t=12 \:\rm ms$ and $\theta=0^\circ$. Last two columns give trap mean frequency $\bar{\omega}$ and anisotropy $\lambda$ for each case.}
\begin{indented}
\item[]
\begin{tabular}{lcllllc}
\br
$^{167}$Er & $A_{\rm K}$ & $A_{\rm R}^{\rm nbal}$ & $A_{\rm R}^{\rm exp}$ & $\Delta A_{\rm R}^{\rm exp}$ & $\bar{\omega} \:\rm (Hz)$ & $\lambda$\\ 
\mr
Case 1 & 1.0258 & 1.0324 & 1.0321 & 0.0012 & 318$\times 2\pi$ &6.54\\
Case 2 & 1.0232 & 1.0282 & 1.0292 & 0.0015 & 261$\times 2\pi$ &4.87\\
Case 3 & 1.0253 & 1.0270 & 1.0258 & 0.0020 & 311$\times 2\pi$ &1.78\\
\br
\end{tabular}
\end{indented}
\end{table}

The discrepancies observed in the figure can be accredited to the effects of the DDI, which are neglected during the TOF by using the ballistic approximation. Even better agreement between the experiment and the theory can be expected if a non-ballistic expansion would be taken into account. Although a theory for this is not yet available for an arbitrary orientation of the dipoles, reference~\cite{Veljic1} allows us to perform a non-ballistic expansion calculation for the special case $\theta=0^\circ$ in the collisionless regime. The comparison of the results is given in table~\ref{tab:tab2}, where we see that accounting for the DDI during the TOF yields theoretical values of the TOF real-space aspect ratio equal to the experimental ones, within the error bars of the order of 0.1\%. Table~\ref{tab:tab2} also shows that non-ballistic effects amount to 0.7\% for Case 1, which has the largest anisotropy, and becomes smaller as the trap is closer to a spherical shape, i.e., as the trap anisotropy approaches the value of 1. Therefore, we conclude that the agreement of experimental data and theoretical results in figure~\ref{fig:comparison}(a) can be further improved by developing a theory for a non-ballistic expansion for a general experiment geometry, which is out of the scope of the present study.

Figure~\ref{fig:comparison}(b) shows a comparison of our theoretical and experimental results for the deformation $\Delta$ of the FS for the three considered cases, where the experimental values are calculated according to (\ref{eq:deltaAk}), assuming ballistic expansion (\ref{eq:AK0}) and using the real-space aspect ratios shown in figure~\ref{fig:comparison}(a). Although the statistical error bars $\Delta A_\mathrm{R}^\mathrm{exp}$ for the experimentally measured values of the real-space aspect ratios are small and almost constant, the corresponding errors for the FS deformation, calculated as
\begin{equation}
\Delta A_\mathrm{R}^\mathrm{exp}\left|\frac{\partial\Delta}{\partial A_\mathrm{K}}\right|_{A_\mathrm{K}=A_\mathrm{R}^\mathrm{exp}}\, ,
\label{eq:err}
\end{equation}
show a strong angular dependence, due to the presence of a pole in the function $\partial\Delta/\partial A_\mathrm{K}$. For the parameters of figure~\ref{fig:comparison}, the pole emerges at around $\theta=50^\circ$. Therefore, the error bars appear significantly larger in the neighbouring region, which justifies to drop the data points around $\theta=50^\circ$ (shaded area in the graph).

As can be seen in figure~\ref{fig:comparison}(b), for all three cases the deformation of the FS is almost constant for all angles $\theta$. Therefore, from the experimental point of view, it would be enough just to measure the aspect ratio for one value of $\theta$, e.g., $\theta=0^\circ$ in order to determine the deformation of the FS. However, this is only true for a weak enough DDI. Nevertheless, even if this is the case, the measurement of the angular dependence of $A_\mathrm{R}$ is an indispensable tool for a full verification of the developed theory, as demonstrated in figure~\ref{fig:comparison}(a) and \ref{fig:comparison}(b).

\subsection{Universal consequences of geometry}
\label{sec:universal}

\begin{figure}[!b]
\centering
\includegraphics[width=8cm]{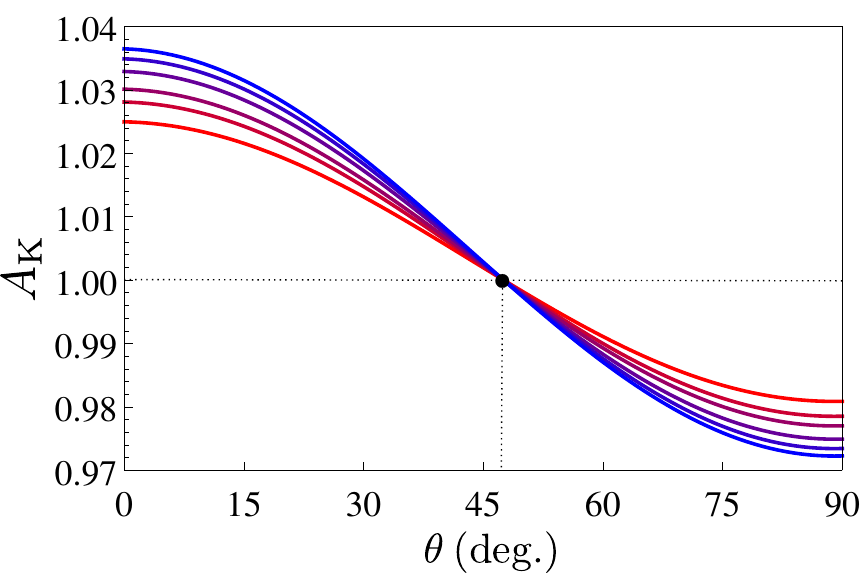}
\caption{$\theta$ dependence of aspect ratio $A_{\rm K}$ for $\rm ^{167}Er$ for $\varphi=0^\circ$, $\alpha=28^\circ$, $N=7 \times 10^4$ and $\omega_x=\omega_z=500\times 2\pi$~Hz. Different curves correspond to varying $\omega_y=n\times 100 \times 2\pi$~Hz, where $n\in\{1,2,3,5,7,9\}$, bottom to top on the left hand side of the intersection, respectively. Intersection point is at ($\theta^*$, $A_\mathrm{K}^*$)=($48.56^\circ$, 1).}
\label{fig:ak_theta}
\end{figure}

As already observed in figure~\ref{fig:comparison}(a), the $A_{\rm K}$ curves for all three considered cases intersect at a special point ($\theta^*$, $A_\mathrm{K}^*=1$). Figure~\ref{fig:ak_theta} reveals that this is not just a coincidence. It shows the $\theta$-dependence of the momentum-space aspect ratio $A_{\rm K}$ for several trapping geometries for erbium atomic gases, ranging from a cigar-shaped trap, through a spherical, to a pancake-shaped trap. The azimuthal angle is kept constant at the value $\varphi=0^\circ$, as well as the trapping frequencies $\omega_x=\omega_z=500\times 2\pi$~Hz, while the frequency $\omega_y=n\times 100 \times 2\pi$~Hz is varied by changing the value $ n\in \{1,2,3,5,7,9\}$, which corresponds to the trap anisotropy $\lambda=5/n$. The number of particles was fixed at $N=7 \times 10^4$. We observe again that all curves intersect for $A_{\rm K}^*=1$, which suggests that this is a general rule.
Indeed, if we take into account that $K_z'\geq K_x'>0$, for $A_{\rm K}^*=1$ we can show from (\ref{eq:Ak}) that the following relation holds, which connects the intersection angles $\theta^*$ and $\varphi^*$:
\begin{equation}
\sin^2\theta^*=\frac{1}{1+\cos^2\varphi^* \cos^2\alpha+\sin^2\varphi^*\sin^2\alpha}\, .
\label{eq:ThetaPhi}
\end{equation}
This result is universal, i.e., it is independent on other system parameters as the trap geometry, the number of particles, and the DDI strength. In other words, this intersection point is purely a consequence of the geometry, and for any orientation of the dipoles there exists an imaging angle such that the aspect ratio is given by $A_\mathrm{K}=1$, while the FS deformation $\Delta$ can be nontrivial and even can have a significant value. We note that for larger $\varepsilon_\mathrm{dd}$ values additional parameter-specific intersection points may appear for some geometries, but the intersection point for $A_\mathrm{K}=1$ is universal and always present.
 
\begin{figure}[!t]
\centering
\includegraphics[width=8cm]{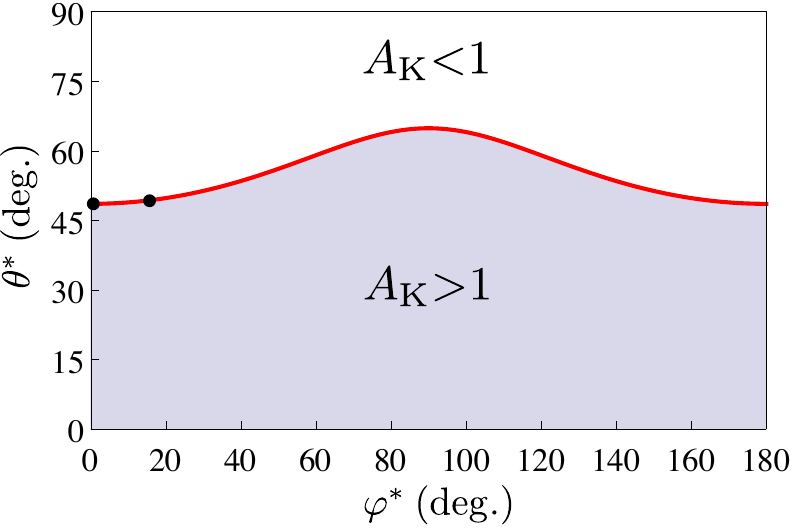}
\caption{Relation between intersection angles $\theta^*$ and $\varphi^*$ for $\alpha=28^\circ$, determined by (\ref{eq:ThetaPhi}): red solid line corresponds to $A_{\rm K}^*=1$, region below to $A_{\rm K} > 1$, and region above to $A_{\rm K} < 1$. Black dots correspond to intersection points identified in figures~\ref{fig:ak_theta} and \ref{fig:comparison}(a) for $\varphi=0^\circ$ and $14^\circ$, respectively.}
\label{fig:ak_theta_phi}
\end{figure}

To further illustrate this, in figure~\ref{fig:ak_theta_phi} we plot a diagram in the $(\theta^*,\, \varphi^*)$-plane for $\alpha=28^\circ$, where the regions with $A_{\rm K} > 1$ and $A_{\rm K} < 1$ are delineated by a solid line defined by (\ref{eq:ThetaPhi}). The two black dots correspond to intersection points from figure~\ref{fig:ak_theta} for $\varphi=0^\circ$ and from figure~\ref{fig:comparison}(a) for $\varphi=14^\circ$, respectively.

\section{Conclusions}
\label{sec:con}

In conclusion, we have explored the ground-state properties of dipolar Fermi gases in elongated traps at zero temperature for an arbitrary orientation of the dipoles. By means of a Hartree-Fock mean-field theory with an appropriate ansatz for the Wigner function, we have shown that the ground-state FS is deformed into an ellipsoid, the main axis of which coincides with the orientation of the dipoles. We have then used this theory to study effects of the dipoles' orientation, the particle number and the trap anisotropy on the deformation of the FS. We have found that the FS deformation is maximal when the dipoles point along the axis with the smallest trapping frequency and demonstrated this for two samples of different dipolar strengths, values of which are achievable with atomic erbium in one case and with polar molecules in the other case. Furthermore, for the erbium case we have observed that the angular dependence of the FS deformation is larger than the corresponding dependence on the trap anisotropy, and that both are less pronounced than the corresponding effect when the number of particles is varied. Note, however, that a stronger DDI may modify this behaviour.

We have established a relationship between the FS deformation and the momentum-space aspect ratio for a general system geometry, which is experimentally accessible by measuring the real-space aspect ratio during the TOF, if we assume ballistic expansion. We have performed new measurements on degenerate gases of atomic $^{167}$Er in different trap geometries and found very good agreement, without any free fitting parameters. As an extension, using pure geometric considerations, we have shown that the FS deformation can have a nontrivial value even when the measured TOF real-space aspect ratio equals one. Furthermore, we have derived a relation between the orientation of the dipoles and the imaging direction for which the aspect ratio is always equal to one.

The theory for the ground-state properties of trapped Fermi gases of tilted dipoles presented here may be relevant for a precise calculation of the critical temperature for BCS pairing of dipolar fermions \cite{Baranov-pairing}. Furthermore, it can also serve as a basis for a further study of the system's dynamics, such as the TOF analysis and the low-lying excitations. Indeed, a proper comparison between theory and experiment for the finite-time real-space aspect ratio, to the best of our knowledge, is still lacking for an arbitrary geometry. These aspects of dipolar Fermi gases, whose understanding is highly relevant for current and future experiments, will be studied in a forthcoming publication. 

\section*{Acknowledgements}
We thank A.~Patscheider and D.~Petter for their help in the experimental measurements and for fruitful discussions.
This work was supported in part by the Ministry of Education, Science and Technological Development of the Republic of Serbia under projects ON171017, BEC-L and DUDFG, by the German Academic and Exchange Service (DAAD) under project BEC-L, by the German Research Foundation (DFG) via the project PE 530/5-1, as well as the Collaborative Research Centers SFB/TR49 and SFB/TR185, and by the Austrian Agency for International Mobility and Cooperation in Education, Science and Research (OeAD) under project DUDFG. A.~R.~P.~L. acknowledges financial support from the Brazilian Funda\c{c}\~ao Cearense de Apoio ao Desenvolvimento Cient\'ifico e Tecnol\'ogico (Grant No.~BP2-0107-00129.02.00/15). The Innsbruck team gratefully acknowledges financial support from the European Research Council through the ERC Consolidator Grant RARE (Grant No.~681432), from the Austrian Science Fund (FWF) project I2790, and from the German Research Foundation (DFG) Research Unit FOR 2247. L.~C.~is supported within the Marie Curie project DipPhase (Grant No.~706809) of the EU H2020 programme. Numerical simulations were run on the PARADOX supercomputing facility at the Scientific Computing Laboratory of the Institute of Physics Belgrade.

\appendix
\section{Hartree energy} 
\label{sec:HE}

The Hartree term (\ref{Edf0}) can be written in terms of the Fourier transform of the potential according to
\begin{equation}
E_{\rm dd}^{\rm D} =\frac{1}{2} \int \frac{d^3k''}{(2\pi)^3} \tilde{V}_{\rm dd}({\bf k''}) \int \frac{d^3k}{(2\pi)^3} \tilde{\nu}({\bf -k''},{\bf k})  \int \frac{d^3k'}{(2\pi)^3} \tilde{\nu}({\bf k''},{\bf k'}). \label{unterteiltEd}
\end{equation}
Defining $h({\bf k})=1-\sum_{ij} k_i  \mathbb{B}_{ij} k_j$ as a suitable abbreviation for the momentum part of the argument of the Wigner function (\ref{eq:ansatzwigner}), the Fourier-transformed distribution function yields
\begin{equation}
\fl
 \tilde{\nu}(-{\bf k}'',{\bf k})=\frac{(2\pi)^{\frac{3}{2}}\bar{R}^3h({\bf k})^{\frac{3}{4}}{\Theta}[h({\bf k})]}{(k_x''^2R_x^2+k_y''^2R_y^2+k_z''^2R_z^2)^{\frac{3}{4}}}J_{\frac{3}{2}}\left[ h({\bf k})^{\frac{1}{2}}\left(k_x''^2R_x^2+k_y''^2R_y^2+k_z''^2R_z^2\right)^{\frac{1}{2}}  \right]\, , \label{FTW1}
\end{equation}
where $J_a$ represents the Bessel function of the first kind.
Then, after an algebraic substitution and switching into spherical coordinates, the integral yields
\begin{eqnarray}
E_{\rm dd}^{\rm D}=&&\frac{\mu_0m^2 \bar{R}^3 \bar{K'}^6}{48\pi^3}\int_0^{\pi}d\vartheta\int_0^{2\pi}d\phi \sin\vartheta \int_0^{\infty}du\, u^{-4}J_{3}^2(u)   \nonumber \\
&&  \times  \left[\frac{3\sin^2\theta\cos^2\varphi\, {\sin}^3\vartheta \cos^2\phi }{{\rm cos}^2\phi{\rm sin}^2\vartheta+(R_x/R_y)^2{\rm sin}^2\phi{\rm sin}^2\vartheta+(R_x/R_z)^2{\rm cos}^2\vartheta} \right.\nonumber\\
&&+ \frac{3\sin^2\theta\sin^2\varphi \,{\sin}^3\vartheta \sin^2\phi }{(R_y/R_x)^2{\rm cos}^2\phi{\rm sin}^2\vartheta+{\rm sin}^2\phi{\rm sin}^2\vartheta+(R_y/R_z)^2{\rm cos}^2\vartheta }   \nonumber \\
&&\left. +\frac{  3\cos^2\theta\,{\rm cos}^2\vartheta \sin\vartheta}{(R_z/R_x)^2{\rm cos}^2\phi{\rm sin}^2\vartheta+(R_z/R_y)^2{\rm sin}^2\phi{\rm sin}^2\vartheta+{\rm cos}^2\vartheta} -1\right].
\end{eqnarray}
Subsequently, we apply \cite[(6.574.2)]{Grad} for the radial integral and make use of the anisotropy function described in \ref{sec:af}, equations~(\ref{Int1})--(\ref{Int3}), so that the Hartree energy $E_{\rm dd}^{\rm D}$ yields 
\begin{eqnarray}
 E_{\rm dd}^{\rm D}=&&-\frac{6N^2 c_0}{ \bar{R}^3}\left[1-3\sin^2\theta\cos^2\varphi f_{A1}\left( \frac{R_x}{R_z},\frac{R_y}{R_z} \right)\right.\nonumber\\
 &&\left. -3\sin^2\theta\sin^2\varphi  f_{A2}\left( \frac{R_x}{R_z},\frac{R_y}{R_z} \right) -3\cos^2\theta f_{A3}\left( \frac{R_x}{R_z},\frac{R_y}{R_z} \right) \right].
\label{eq:AEdd}
\end{eqnarray}
Finally, the identities (\ref{identitiesofanisotropyfunction}) lead to
\begin{equation}
 E_{\rm dd}^{\rm D}=-\frac{6N^2 c_0}{ \bar{R}^3}f_{\rm A}\left( \frac{R_x}{R_z},\frac{R_y}{R_z},\theta,\varphi\right),
\label{Ehartree}
\end{equation}
with the definition (\ref{gen_aniso_func}) for the generalised anisotropy function.

\section{Fock energy}
\label{sec:FE}

The Fock term can be rewritten in the following form
\begin{equation}
\fl
E_{\rm dd}^{\rm E} =-\frac{1}{2}\int d^3x' \int \frac{d^3k'}{(2\pi)^3}\int \frac{d^3k''}{(2\pi)^3} \bar{\tilde{\nu}}({\bf k''},{\bf x'}) \bar{\tilde{\nu}}(-{\bf k''},-{\bf x'})\tilde{V}_{\rm dd}({\bf k'}) {\rm e}^{i{\bf x'}\cdot {\bf k'}}\, ,  \label{generalfockequation}
\end{equation}
where $\bar{\nu}({\bf k'},{\bf k})$ and $\tilde{\nu}({\bf x},{\bf x'})$ denote the Fourier transforms of $\nu({\bf x},{\bf k})$ with respect to the first and second variable, respectively. The first step is to calculate these two Fourier transforms of the Wigner function. The first of these two transforms has already been obtained in (\ref{FTW1}). Using this result, one gets
\small
\begin{eqnarray}
\fl
 \bar{\tilde{\nu}}(-{\bf k}'',{\bf x})=\int \frac{d^3k}{(2\pi)^3}{\rm e}^{i {\bf k}\cdot {\bf x}}\tilde{\nu}(-{\bf k}'',{\bf k})=
\int \frac{d^3q}{(2\pi)^3}{\rm e}^{i {\bf q}\cdot (\mathbb{R}^T{\bf x})}\tilde{\nu}(-{\bf k}'',{\mathbb{R} \bf q}) \nonumber \\ 
\fl
=\int \frac{d^3q}{(2\pi)^{\frac{3}{2}}} {\rm e}^{i{\bf q} \cdot {\bf c}} \frac{\bar{R}^3\Theta\left(1-\sum_j \frac{q_j^2}{K_j'^2}\right)}{g({\bf k}'')^{\frac{3}{4}}}^{\frac{3}{4}}\left( 1-\sum_l \frac{q_l^2}{K_l'^{2}} \right)^{\frac{3}{4}} J_{\frac{3}{2}}\left[ \left(1-\sum_m \frac{q_m^2}{K_m'^{2}} \right)^{\frac{1}{2}} g({\bf k}'')^{\frac{1}{2}}\right], \label{secondfouriertrafo}
\end{eqnarray}
\normalsize
where $g({\bf k}'')=k_x''^2R_x^2+k_y''^2R_y^2+k_z''^2R_z^2$ and ${\bf c}=\mathbb{R}^T{\bf x}$. The three $q$-integrals will be treated separately, yet all in the same way. Hence, it is only necessary to compute one of them. With the use of the substitution $q_z=K_z'\sqrt{1-\frac{q_x^2}{K_x'^2}-\frac{q_y^2}{K_y'^2}}\cos \vartheta$, one can rewrite (\ref{secondfouriertrafo}) as
\begin{eqnarray}
\fl
 \bar{\tilde{\nu}}(-{\bf k}'',{\bf x})=\frac{\bar{R}^3}{(2\pi)^{\frac{3}{2}}}\frac{2}{g(\bf k)''^{\frac{3}{4}}} \int dq_x dq_y {\rm e}^{ic_xq_x+ic_yq_y}\Theta\left( 1-\frac{q_x^2}{K_x'^{2}}-\frac{q_y^2}{K_y'^{2}}\right) \left( 1-\frac{q_x^2}{K_x'^{2}}-\frac{q_y^2}{K_y'^{2}}\right)^{\frac{5}{4}}  \nonumber \\
\times \int_0^{\frac{\pi}{2}} d\vartheta \sin^{\frac{5}{2}}\vartheta K_z' \cos\left( c_zK_z' \sqrt{1-\frac{q_x^2}{K_x'^{2}}-\frac{q_y^2}{K_y'^{2}}}\cos \vartheta \right) \nonumber\\
\times J_{\frac{3}{2}}\left[ g({\bf k}'')^{\frac{1}{2}} \left( 1-\frac{q_x^2}{K_x'^{2}}-\frac{q_y^2}{K_y'^{2}} \right)^{\frac{1}{2}} \sin \vartheta \right].
\end{eqnarray}
After this substitution, the $\vartheta$-integral can be calculated using \cite[(6.688.2)]{Grad}. The other two $k$-integrals can be treated in the same way. Then, the Fourier transform reads 
\begin{equation}
\fl
 \bar{\tilde{\nu}}(-{\bf k}'',{\bf x})=\frac{\bar{R}^3 \bar{K'}^3}{\left[g({\bf k}'')+c_z^2K_z'^2+c_y^2K_y'^2+c_x^2K_x'^2\right]^{\frac{3}{2}}}J_3\left\{ \left[g({\bf k}'')+c_z^2K_z'^2+c_y^2K_y'^2+c_x^2K_x'^2\right]^{\frac{1}{2}} \right\}.
\end{equation}
It is clear that $\bar{\tilde{\nu}}({\bf k}'',{\bf x})$ is an even function, which further simplifies the calculations. The next step is to calculate the ${\bf x}'$-integral in (\ref{generalfockequation}). To avoid a quadratic Bessel function, one can use the integral representation \cite[(6.519.2.2)]{Grad}, which leads to an integral over a Bessel function 
\begin{eqnarray}
 J_3^2\left\{\left[c_x^2K_x'^2+c_y^2K_y'^2+c_z^2K_z'^2+g({\bf k}'')\right]^{\frac{1}{2}}\right\}\nonumber\\
 =\frac{2}{\pi}\int_0^{\frac{\pi}{2}} dt J_6 \left\{ 2\sin t \left[ c_x^2K_x'^2+c_y^2K_y'^2+c_z^2K_z'^2+g({\bf k}'') \right]^{\frac{1}{2}} \right\} \label{Besselidentity}.
\end{eqnarray}
With this the ${\bf x}'$-integral becomes
\begin{eqnarray*}
\fl
\int d^3x' \bar{\tilde{\nu}}({\bf k}'',{\bf x'})^2 {\rm e}^{i{\bf k'}\cdot {\bf x'}}= 
\int d^3x'' \bar{\tilde{\nu}}({\bf k}'',{\mathbb{R} \bf x''})^2 {\rm e}^{i{\bf k'}\cdot (\mathbb{R}{\bf x''})} =
\int d^3x'' \bar{\tilde{\nu}}({\bf k}'',{\mathbb{R} \bf x''})^2 {\rm e}^{i (\mathbb{R}^T{\bf k'})^T\cdot {\bf x''}} \nonumber \\
\fl
 =\int d^3x'' \frac{\bar{R}^6\bar{K'}^6  {\rm e}^{i {\boldsymbol \kappa} \cdot {\bf x''}}}{\left[ x''^2 K_x'^2+y''^2K_y'^2+ z''^2K_z'^2+g({\bf k}'') \right]^3} J_3^2\left\{  \left[ x''^2 K_x'^2+y''^2K_y'^2+ z''^2K_z'^2+g({\bf k}'') \right]^{\frac{1}{2}} \right\}\nonumber \\
 \fl
 = \int d^3x'' \frac{\bar{R}^6\bar{K'}^6  {\rm e}^{i {\boldsymbol \kappa} \cdot {\bf x''}}}{\left[ x''^2 K_x'^2+y''^2K_y'^2+ z''^2K_z'^2+g({\bf k}'') \right]^3}\nonumber\\
 \times \frac{2}{\pi}\int_0^{\frac{\pi}{2}} dt J_6 \left\{ 2\sin t \left[  x''^2K_x'^2+ y''^2K_y'^2+z''^2K_z'^2+g({\bf k}'') \right]^{\frac{1}{2}} \right\},
\end{eqnarray*}
\normalsize
where ${\boldsymbol \kappa}=\mathbb{R}^T{\bf k'}$. We treat the three integrals separately, starting with the $z''$-integral. After the substitution $u_z=z''K_z'$, we can use \cite[(6.726.2)]{Grad} to evaluate the integral over $z''$
\begin{eqnarray}
\fl
 \bar{R}^6\bar{K'}^6\int dx''dy'' {\rm e}^{i\kappa_x x''+i\kappa_y y''}\frac{4}{\pi K_z'}\int_0^{\frac{\pi}{2}}dt \frac{\left(4\sin ^2t-\frac{\kappa_z^2}{K_z'^2}\right)^{\frac{11}{4}}\Theta\left( 2\sin t-\sqrt{\frac{\kappa_z^2}{K_z'^2}} \right)}{(2 \sin t)^6\left[ x''^2 K_x'^2+y''^2K_y'^2+g({\bf k}'') \right]^{\frac{11}{4}}} \nonumber \\
\times \sqrt{\frac{\pi}{2}} J_{\frac{11}{2}}\left\{ \left( 4\sin ^2t-\frac{\kappa_z^2}{K_z'^2} \right)^{\frac{1}{2}} \left[ x''^2 K_x'^2+y''^2K_y'^2+g({\bf k}'') \right]^{\frac{1}{2}} \right\}.
\end{eqnarray}
The other two integrals will be calculated in the same way. The solution of the ${\bf x}'$-integral reads
\begin{eqnarray}
\fl
 \int d^3x' \bar{\tilde{\nu}}({\bf k}'',{\bf x'})^2 {\rm e}^{i{\bf k'}\cdot {\bf x'}}=\left(\frac{\pi}{2^7}\right)^{\frac{1}{2}}\bar{R}^6\bar{K'}^3\int_0^{\frac{\pi}{2}}\frac{dt}{\sin^6t}\frac{\left( 4\sin^2 t -\frac{\kappa_z^2}{K_z'^2}-\frac{\kappa_y^2}{K_y'^2}-\frac{\kappa_x^2}{K_x'^2}\right)^{\frac{9}{4}}}{g({\bf k}'')^{\frac{9}{4}}} \nonumber \\
 \fl
\times J_{\frac{9}{2}}\left[ g({\bf k}'')^{\frac{1}{2}}\left( 4\sin^2 t -\frac{\kappa_z^2}{K_z'^2}-\frac{\kappa_y^2}{K_y'^2}-\frac{\kappa_x^2}{K_x'^2}\right)^{\frac{1}{2}} \right] \Theta\left(2\sin t-\sqrt{\frac{\kappa_z^2}{K_z'^2}+\frac{\kappa_y^2}{K_y'^2}+\frac{\kappa_x^2}{K_x'^2}}\, \right).
\end{eqnarray}
The next step is to integrate the $\bf{k''}$-integral. Using the underlying spherical symmetry, the calculation of this integral becomes relatively short. Indeed, substituting $u_i=k_i''R_i$ and then transforming these new integration variables into spherical coordinates, one can use \cite[(6.561.17)]{Grad}, which leads to
\begin{eqnarray}
\fl
 I({\bf k}')=\int d^3k''\int d^3x' \bar{\tilde{\nu}}({\bf k}'',{\bf x'})^2 {\rm e}^{i{\bf k'}\cdot {\bf x'}}=\frac{\pi^2\bar{R}^3\bar{K'}^3}{192}\int_0^{\frac{\pi}{2}}\frac{dt}{\sin^6t}\nonumber \\
 \hspace*{-8mm}
 \times\left( 4\sin^2 t -\frac{\kappa_z^2}{K_z'^2}-\frac{\kappa_y^2}{K_y'^2}-\frac{\kappa_x^2}{K_x'^2}\right)^3\,
 \Theta\left(2\sin t-\sqrt{\frac{\kappa_z^2}{K_z'^2}+\frac{\kappa_y^2}{K_y'^2}+\frac{\kappa_x^2}{K_x'^2}}\,\right). \label{doubleintdoublefourexp}
\end{eqnarray}
The last step of the calculation of the Fock term is to solve the $k'$-integral, and therefore we will again switch to another coordinate system ${\bf k'}=\mathbb{R} {\bf q}$ and use the Fourier transform of the interaction potential
\begin{eqnarray}
E_{\rm dd}^{\rm E}&=&-\frac{1}{2 (2\pi)^6} \int d^3k' I({\bf k}')\tilde{V}_{\rm dd}({\bf k'})=
-\frac{1}{2 (2\pi)^6}\int d^3q  I({\mathbb{R} \bf q}) \tilde{V}_{\rm dd}({{\mathbb{R} \bf q}}) \nonumber\\
&=&-\frac{\mu_0m^2 \bar{R}^3\bar{K'}^3}{73728\pi^4} \int d^3q \int_0^{\frac{\pi}{2}}\frac{dt}{\sin^6t}
\left( 4\sin^2 t -\frac{q_z^2}{K_z'^2}-\frac{q_y^2}{K_y'^2}-\frac{q_x^2}{K_x'^2}\right)^3\nonumber\\
&&\times \Theta\left(2\sin t-\sqrt{\frac{q_z^2}{K_z'^2}+\frac{q_y^2}{K_y'^2}+\frac{q_x^2}{K_x'^2}}\right)\left( \frac{3q_z^2}{q^2}-1\right). \label{doubleintdoublefourexp1}
\end{eqnarray}
Using the substitution $u_i=q_i/K_i'$ and afterwards switching to spherical coordinates we get 
\begin{eqnarray}
\hspace*{-12mm}
E_{\rm dd}^{\rm E}=-\frac{\mu_0m^2 \bar{R}^3\bar{K'}^6}{73728\pi^4}\int_0^{2\pi}d\phi \int_0^{\pi} d\vartheta  \int_0^{\frac{\pi}{2}}\frac{dt}{\sin^6t} \int_0^{2 \sin t}du u^2 \left( 4\sin ^2 t-u^2 \right)^3  \nonumber \\
 \times  \left[ 3\frac{{\rm cos}^2\vartheta \sin\vartheta}{(K_x'/K_z')^2{\rm cos}^2\phi{\rm sin}^2\vartheta+(K_y'/K_z')^2{\rm sin}^2\phi{\rm sin}^2\vartheta+{\rm cos}^2\vartheta} -\sin\vartheta\right].
\end{eqnarray}
The $\vartheta$- and $\phi$-integrals will lead to the anisotropy function, which is defined by (\ref{Int3}) and (\ref{labelC6}), and the $u$-integral and $t$-integral can be solved without any difficulties. The solution of the Fock term reads
\begin{equation}
 E_{\rm dd}^{\rm E}=\frac{6N^2c_0}{\bar{R}^3}\left[1-3f_{A3}\left( \frac{K_z'}{K_x'},\frac{K_z'}{K_y'} \right)\right]=\frac{6N^2c_0}{\bar{R}^3}f\left( \frac{K_z'}{K_x'},\frac{K_z'}{K_y'} \right).
\end{equation}

\section{Anisotropy function} 
\label{sec:af}
Here we recall the definition of the usual anisotropy function \cite{Falk}
\begin{eqnarray}
\hspace*{-15mm}
f(x,y) & = & -\frac{1}{4\pi}\int_0^{2\pi}d\phi \int_0^{\pi}d\vartheta \, {\rm sin}\vartheta \left[ \frac{3x^2y^2{\rm cos}^2\vartheta}{\left(y^2{\rm cos}^2 
\phi+x^2{\rm sin}^2\phi\right) {\rm sin}^2\vartheta+x^2y^2{\rm cos}^2\vartheta}-1\right]\nonumber\\
\hspace*{-15mm}
& = & 1 + 3 x y \, \frac{E(\vartheta_x,\kappa) - F(\vartheta_x,\kappa)}{(1-y^2)\sqrt{1-x^2}} \, ,
\label{definitionanistropyfunction}
\end{eqnarray}
where $\vartheta_x=\arccos x$, $\vartheta_y=\arccos y$, $\kappa^2=(1-y^2)/(1-x^2)$ are abbreviations and, $F(\varphi,k)$ is the elliptic integral of first kind and $E(\varphi,k)$ is the elliptic integral of second kind.
Notice that
\begin{equation}
f(x,x)=f_{s}(x)
\end{equation}
denotes the cylindrically symmetric anisotropy function \cite{yi_s_2001,odell_dhj_2004,Glaum1,Glaum2}. We, then, introduce some auxiliary functions
\begin{eqnarray}
\hspace*{-10mm}
4\pi f_{A1}(x,y)&=&\int_0^{2\pi}d\phi \int_0^{\pi}d\vartheta \frac{y^2{\sin}^3\vartheta \cos^2\phi }{y^2{\rm cos}^2\phi{\rm sin}^2\vartheta+x^2{\rm sin}^2\phi{\rm sin}^2\vartheta+x^2y^2{\rm cos}^2\vartheta}\nonumber\\
\hspace*{-10mm}&=&4\pi \frac{y^2}{y^2-x^2}\left[1-\frac{x}{y}\frac{E(\vartheta_x,\kappa)}{\sqrt{1-x^2}} \right]\, , \label{Int1} \\ 
\hspace*{-10mm}
4\pi f_{A2}(x,y)&=&\int_0^{2\pi}d\phi \int_0^{\pi}d\vartheta \frac{x^2{\sin}^3\vartheta \sin^2\phi }{y^2{\rm cos}^2\phi{\rm sin}^2\vartheta+x^2{\rm sin}^2\phi{\rm sin}^2\vartheta+x^2y^2{\rm cos}^2\vartheta}\nonumber\\
\hspace*{-10mm}&=&4\pi\frac{ x^2}{x^2-y^2}\left[ 1-\frac{y}{x}\frac{E(\vartheta_y,\frac{1}{\kappa})}{\sqrt{1-y^2}}\right]\, , \label{Int2} \\  
\hspace*{-10mm}
4\pi f_{A3}(x,y)&=&\int_0^{2\pi}d\phi \int_0^{\pi}d\vartheta \,   \frac{x^2y^2{\rm cos}^2\vartheta {\rm sin}\vartheta}{y^2{\rm cos}^2 \phi {\rm sin}^2\vartheta+x^2{\rm sin}^2\phi {\rm sin}^2\vartheta+x^2y^2{\rm cos}^2\vartheta}\nonumber\\
\hspace*{-15mm}&=&-4\pi xy\frac{E(\vartheta_x,\kappa)-F(\vartheta_x,\kappa)}{(1-y^2)\sqrt{1-x^2}}\, . \label{Int3}
\end{eqnarray}

At this point, the following identities can be derived
\begin{eqnarray}
f(x,y)&=1-3f_{A3}(x,y)\, , \label{labelC6}\\
f_{A1}(x,y)&= f_{A3}(y/x,1/x)\, ,  \\
f_{A2}(x,y)&=f_{A3}(x/y,1/y)\, , \label{labelC8}\\
\sum_{i=1,2,3} f_{Ai}(x,y)&=1\, .
\label{identitiesofanisotropyfunction}
\end{eqnarray}
Thus, finally, taking (\ref{labelC6})--(\ref{labelC8}) into account, we have
\begin{eqnarray}
\hspace*{-25mm}
f_{\rm A}\left(x,y,\theta,\varphi\right) &=&1-3\sin^2\theta\cos^2\varphi\, f_{A1}(x,y) -3\sin^2\theta\sin^2\varphi\, f_{A2}(x,y) -3\cos^2\theta\,  f_{A3}(x,y) \nonumber\\
\hspace*{-25mm}
& = & \sin^2\theta\cos^2\varphi\,  f\left( \frac{y}{x},\frac{1}{x} \right)  +\sin^2\theta\sin^2\varphi \,  f\left( \frac{x}{y},\frac{1}{y} \right)  +\cos^2\theta\, f\left( {x},{y} \right)\, .
\end{eqnarray}

\section{Equations for variational parameters in momentum and real space}
\label{var_param}

Here we present the respective equations for the variational parameters of the Wigner function ans\"atze, i.e., Thomas-Fermi radii and momenta, for all three considered scenarios in section~\ref{sec:theo_mod}, see figure~\ref{fig:schemas}.

\subsection{Spherical scenario}

The five variational parameters ($K_{\rm F}$, $R_i$, $\mu$) are determined by minimising the grand-canonical potential $\Omega^{(1)}$, which leads to the following set of algebraic equations:
\small
\begin{eqnarray}
\fl\mu=\frac{\hbar^2K_{\rm F}^2}{8M}\, ,\label{VSTKfa}\\
\fl\omega_x^2R_x^2+\frac{48Nc_0}{M\bar{R}^3}\left[f_{\rm A}\left(\frac{R_x}{R_z},\frac{R_y}{R_z},\theta,\varphi\right) -R_x\partial_{R_x}f_{\rm A}\left(\frac{R_x}{R_z},\frac{R_y}{R_z},\theta,\varphi\right)\right]
-\frac{8\mu}{M}=0\, , \label{VSTRxa} \\
\fl \omega_y^2R_y^2+\frac{48Nc_0}{M\bar{R}^3} \left[f_{\rm A}\left(\frac{R_x}{R_z},\frac{R_y}{R_z},\theta,\varphi\right) -R_y\partial_{R_y}f_{\rm A}\left(\frac{R_x}{R_z},\frac{R_y}{R_z},\theta,\varphi\right) \right] -\frac{8\mu}{M}=0\, , \label{VSTRya} \\
\fl \omega_z^2R_z^2+\frac{48Nc_0}{M\bar{R}^3} \left[f_{\rm A}\left(\frac{R_x}{R_z},\frac{R_y}{R_z},\theta,\varphi\right) -R_z\partial_{R_z}f_{\rm A}\left(\frac{R_x}{R_z},\frac{R_y}{R_z},\theta,\varphi\right)\right] -\frac{8\mu}{M} =0\, , \label{VSTRza}\\
\fl N=\frac{1}{48}\bar{R}^3K_{\rm F}^3\, .\label{VSTNa}
\end{eqnarray}
\normalsize
Note that (\ref{VSTNa}) represents the particle-number conservation constraint, which is the special case of (\ref{eq:partconser}) for $K_x=K_y=K_z=K_\mathrm{F}$.

\subsection{On-axis scenario}
The seven variational parameters ($K_i$, $R_i$, $\mu$) are determined by minimising the grand-canonical potential $\Omega^{(2)}$, which leads to the following set of algebraic equations:
\small
\begin{eqnarray}
\fl\frac{\hbar^2 K_x^2}{2M} +\frac{24 N c_0}{\bar{R}^3}{K_x}\partial_{K_x} f_{A}\left( \frac{K_z}{K_x},\frac{K_z}{K_y},\theta,\varphi\right)-4\mu=0\, , \label{VSTKxb}\\
\fl\frac{\hbar^2 K_y^2}{2M}+ \frac{24 N c_0}{\bar{R}^3} {K_y}\partial_{K_y} f_{A}\left( \frac{K_z}{K_x},\frac{K_z}{K_y},\theta,\varphi\right)-4\mu=0\, , \label{VSTKyb}\\
\fl\frac{\hbar^2 K_z^2}{2M}+ \frac{24 N c_0}{\bar{R}^3}{K_z}\partial_{K_z} f_{A}\left( \frac{K_z}{K_x},\frac{K_z}{K_y},\theta,\varphi \right)-4\mu=0\, , \label{VSTKzb} \\
\fl\omega_x^2R_x^2+\frac{48Nc_0}{M\bar{R}^3}\left[f_{\rm A}\left(\frac{R_x}{R_z},\frac{R_y}{R_z},\theta,\varphi\right) -f_{A}\left( \frac{K_z}{K_x},\frac{K_z}{K_y},\theta,\varphi \right)-R_x\partial_{R_x}f_{\rm A}\left(\frac{R_x}{R_z},\frac{R_y}{R_z},\theta,\varphi\right)\right]-\frac{8\mu}{M}=0\, , \label{VSTRxb} \\
\fl\omega_y^2R_y^2+\frac{48Nc_0}{M\bar{R}^3} \left[f_{\rm A}\left(\frac{R_x}{R_z},\frac{R_y}{R_z},\theta,\varphi\right) -f_{A}\left( \frac{K_z}{K_x},\frac{K_z}{K_y},\theta,\varphi \right)-R_y\partial_{R_y}f_{\rm A}\left(\frac{R_x}{R_z},\frac{R_y}{R_z},\theta,\varphi\right) \right]-\frac{8\mu}{M}=0\, , \label{VSTRyb} \\
\fl
\omega_z^2R_z^2+\frac{48Nc_0}{M\bar{R}^3} \left[f_{\rm A}\left(\frac{R_x}{R_z},\frac{R_y}{R_z},\theta,\varphi\right) -f_{A}\left( \frac{K_z}{K_x},\frac{K_z}{K_y},\theta,\varphi \right)-R_z\partial_{R_z}f_{\rm A}\left(\frac{R_x}{R_z},\frac{R_y}{R_z},\theta,\varphi\right)\right]-\frac{8\mu}{M}=0\, , \label{VSTRzb}\hspace*{11mm}\\
\fl N=\frac{1}{48}\bar{R}^3\bar K^3\, .\label{VSTNb}
\end{eqnarray}
\normalsize
Similarly as in the spherical scenario, (\ref{VSTNb}) coincides with the particle-number conservation equation~(\ref{eq:partconser}).
 
\subsection{Off-axis scenario}
The seven variational parameters ($K'_i$, $R_i$, $\mu$) are determined by minimising the grand-canonical potential $\Omega^{(3)}$, which leads to the following set of algebraic equations:
\small
\begin{eqnarray}
\fl\frac{\hbar^2 K_x'^2}{2M} +\frac{24 N c_0}{\bar{R}^3}{K'_x}\partial_{K'_x} f\left( \frac{K'_z}{K'_x},\frac{K'_z}{K'_y}\right)-4\mu=0\, , \label{VSTKx}\\
\fl\frac{\hbar^2 K_y'^2}{2M}+ \frac{24 N c_0}{\bar{R}^3} {K'_y}\partial_{K'_y} f\left( \frac{K'_z}{K'_x},\frac{K'_z}{K'_y}\right)-4\mu=0\, , \label{VSTKy}\\
\fl\frac{\hbar^2 K_z'^2}{2M}+ \frac{24 N c_0}{\bar{R}^3}{K'_z}\partial_{K'_z} f\left( \frac{K'_z}{K'_x},\frac{K'_z}{K'_y}\right)-4\mu=0\, , \label{VSTKz} \\
\fl\omega_x^2R_x^2+\frac{48Nc_0}{M\bar{R}^3}\left[f_{\rm A}\left(\frac{R_x}{R_z},\frac{R_y}{R_z},\theta,\varphi\right) -f\left( \frac{K'_z}{K'_x},\frac{K'_z}{K'_y}\right)-R_x\partial_{R_x}f_{\rm A}\left(\frac{R_x}{R_z},\frac{R_y}{R_z},\theta,\varphi\right)\right]-\frac{8\mu}{M}=0\, , \label{VSTRx} \\
\fl\omega_y^2R_y^2+\frac{48Nc_0}{M\bar{R}^3} \left[f_{\rm A}\left(\frac{R_x}{R_z},\frac{R_y}{R_z},\theta,\varphi\right) -f\left( \frac{K'_z}{K'_x},\frac{K'_z}{K'_y}\right)-R_y\partial_{R_y}f_{\rm A}\left(\frac{R_x}{R_z},\frac{R_y}{R_z},\theta,\varphi\right) \right]-\frac{8\mu}{M}=0\, , \label{VSTRy} \\
\fl\omega_z^2R_z^2+\frac{48Nc_0}{M\bar{R}^3} \left[f_{\rm A}\left(\frac{R_x}{R_z},\frac{R_y}{R_z},\theta,\varphi\right) -f\left( \frac{K'_z}{K'_x},\frac{K'_z}{K'_y}\right)-R_z\partial_{R_z}f_{\rm A}\left(\frac{R_x}{R_z},\frac{R_y}{R_z},\theta,\varphi\right)\right]-\frac{8\mu}{M}=0\, , \label{VSTRz}\hspace*{11mm}\\
\fl N=\frac{1}{48}\bar{R}^3\bar K'^3\, .\label{VSTN}
\end{eqnarray}
\normalsize
As before, (\ref{VSTN}) coincides with the particle-number conservation equation~(\ref{eq:partconser}).

Due to the symmetry of the anisotropy function $f(x,y)=f(y,x)$, it follows from (\ref{VSTKx}) and (\ref{VSTKy}) that $K'_x=K'_y$, i.e., that the FS is cylindrically symmetric with respect to the dipoles' orientation. Additionally, in close analogy with the special case when the dipoles are aligned with one of the trapping axes \cite{lima1, lima2, Falk,Veljic1}, the three equations~(\ref{VSTKx})--(\ref{VSTKz}) can be rewritten in the following form:
\begin{eqnarray}
&K'_x=K'_y\, , \label{KxKy}  \\
&K_z'^2- K_x'^2=\frac{144MNc_0}{\hbar^2 \bar{R}^3}\left[ 1+\frac{ \left( 2 K_x'^2+ K_z'^2 \right) f_{\rm s}\left( \frac{K_z'}{K_x'} \right)}{2\left(  {K_z'}^2-{ K_x'}^2 \right)} \right]\, ,
\label{cylindersymmglobalmoment}\\
&\mu=\frac{1}{12}\sum_j\frac{\hbar^2K_j'^2}{2M}\, .\label{mu3}
\end{eqnarray}

\end{document}